\documentclass[journal]{IEEEtran}

\usepackage{graphicx}
\usepackage{amsmath}
\usepackage{amssymb}
\usepackage{tabularx}
\usepackage{graphicx}
\usepackage{subcaption}
\usepackage{times}
\usepackage{multicol}
\usepackage{adjustbox}
\usepackage{xcolor}
\usepackage{url} 

\begin{document}
%
\title{Rethinking Theoretical Illumination for Efficient Low-Light Image Enhancement}
%
%
%
\author{
    Shyang-En~Weng,
    Cheng-Yen~Hsiao,
    Li-Wei Lu, Yu-Shen Huang, Tzu-Han Chen,
    ~Shaou-Gang~Miaou,
    and Ricky Christanto
    \thanks{Shyang-En Weng was with the Department of Electronic Engineering, Chung Yuan Christian University, Taoyuan 320, Taiwan, and the Department of Electrical Engineering, University of Wisconsin–Milwaukee, Milwaukee, WI 53211 USA. He is currently with the Institute of Computer Science and Engineering, National Yang Ming Chiao Tung University, Hsinchu 300, Taiwan.}%
    \thanks{Shaou-Gang Miaou, Cheng-Yen Hsiao, Li-Wei Lu, Yu-Shen Huang, Tzu-Han Chen, and Ricky Christanto are with the Department of Electronic Engineering, Chung Yuan Christian University, Taoyuan 320, Taiwan.}
    \thanks{\textit{Corresponding author: Shaou-Gang Miaou} (e-mail:miaou@cycu.edu.tw).}
    
}

\maketitle

\begin{abstract}
Enhancing low-light images remains a critical challenge in computer vision, as does designing lightweight models for edge devices that can handle the computational demands of deep learning. This article introduces an extended version of the Channel-Prior and Gamma-Estimation Network (CPGA-Net), termed CPGA-Net+, incorporating the theoretically-based Attentions for illumination in local and global processing. Additionally, we assess our approach through a theoretical analysis of the block design by introducing both an ultra-lightweight and a stronger version, following the same design principles. The lightweight version significantly reduces computational costs by over two-thirds by utilizing the local branch as an auxiliary component. Meanwhile, the stronger version achieves an impressive balance by maximizing local and global processing capabilities. Our proposed methods have been validated as effective compared to recent lightweight approaches, offering superior performance and scalable solutions with limited computational resources. The source code will be released at \url{https://github.com/Shyandram/CPGA-Net_Plus}.
\end{abstract}


\section{Introduction}
%
%
%
%
\IEEEPARstart{W}{hether}  indoors or outdoors, low-light image capture poses significant challenges for accurate visual analysis. The limited light reflection often results in degraded image quality, including color inaccuracies and increased noise levels. These issues can significantly affect the performance and reliability of light-sensitive applications, such as transportation surveillance and Advanced Driver Assistance Systems. Therefore, it is crucial to address these challenges to ensure the effective operation of systems under low-light conditions.

The problems of low-light image enhancement (LLIE) are commonly addressed using two primary methods: Histogram Equalization~\cite{gonzalez2017digital} and Retinex~\cite{land1977retinex}. Histogram Equalization works by enhancing contrast through the redistribution of grayscale values. On the other hand, the Retinex theory divides the image into reflectance and illumination components to improve reflectance and overall image quality. Techniques such as Single Scale Retinex~\cite{Jobson1997Properties} and Multi-Scale Retinex~\cite{Rahman1996Multi} are particularly effective in preserving details and managing complex lighting conditions.

With technological advancements, various deep learning-based methods~\cite{Chen2018Retinex, Jiang2021EnlightenGAN, zhang2019kind} have been proposed to enhance the quality of low-light images. However, these methods often require substantial computational resources, which limits their practical application on real-world devices. Therefore, designing lightweight and efficient image enhancement techniques is crucial. In our previous study, we introduced the CPGA-Net~\cite{weng2024lightweight}, which combines Retinex theory with the Atmospheric Scattering Model (ATSM)~\cite{mccartney_optics_1976} and utilizes gamma correction for both global and local processing; it highlights the importance of gamma correction in LLIE. CPGA-DIA~\cite{weng2024dia} explores exposure correction and LLIE issues through dynamic gamma adjustment, showing that gamma correction can be efficient and effective for the enhancement process even in deep learning frameworks.

\begin{figure}[t]
    \centering
    \begin{subfigure}[b]{.49\linewidth}
        \centering
        \includegraphics[width=\linewidth]{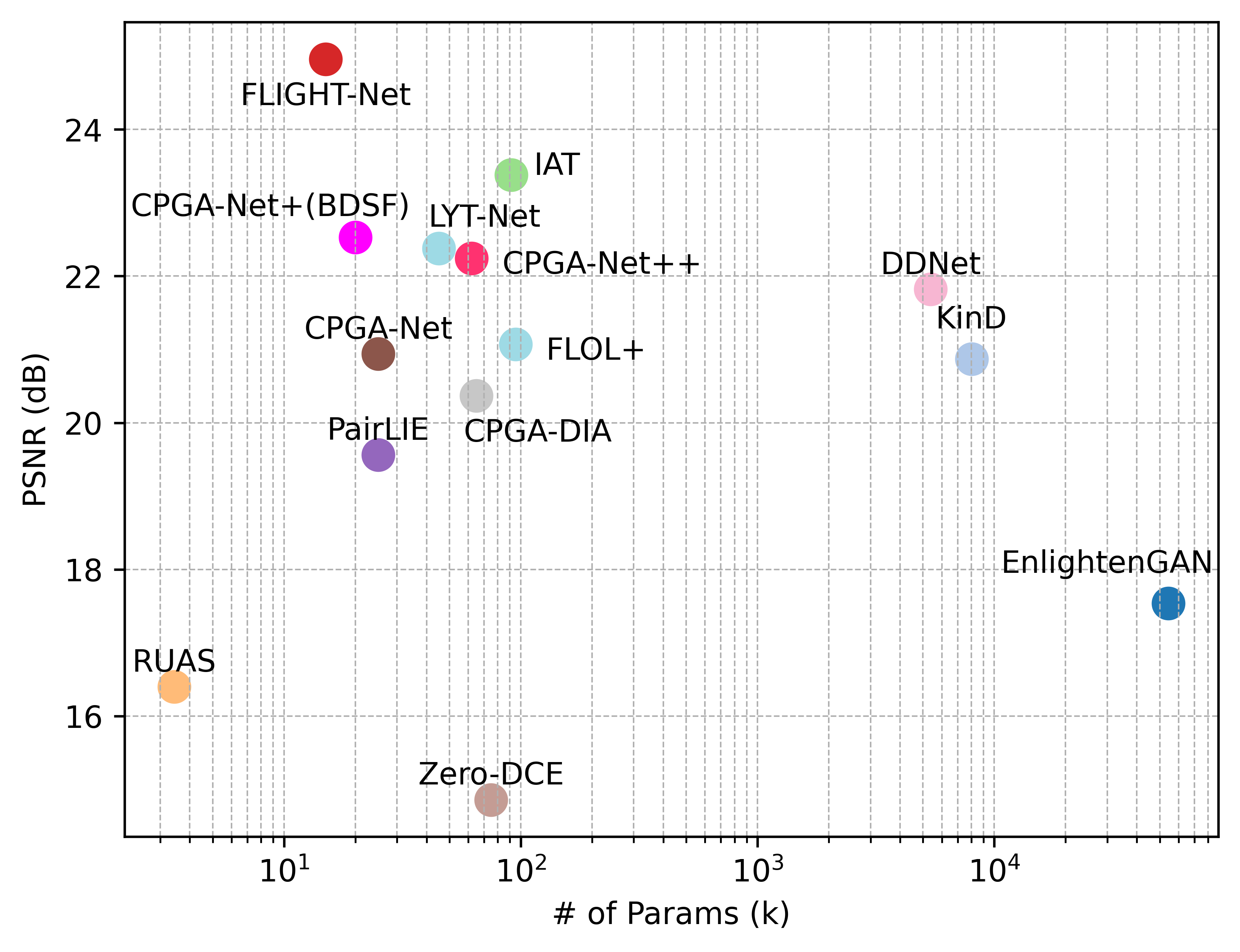}
        \caption{}
     \end{subfigure}
     \begin{subfigure}[b]{.49\linewidth}
        \centering
         \includegraphics[width=\linewidth]{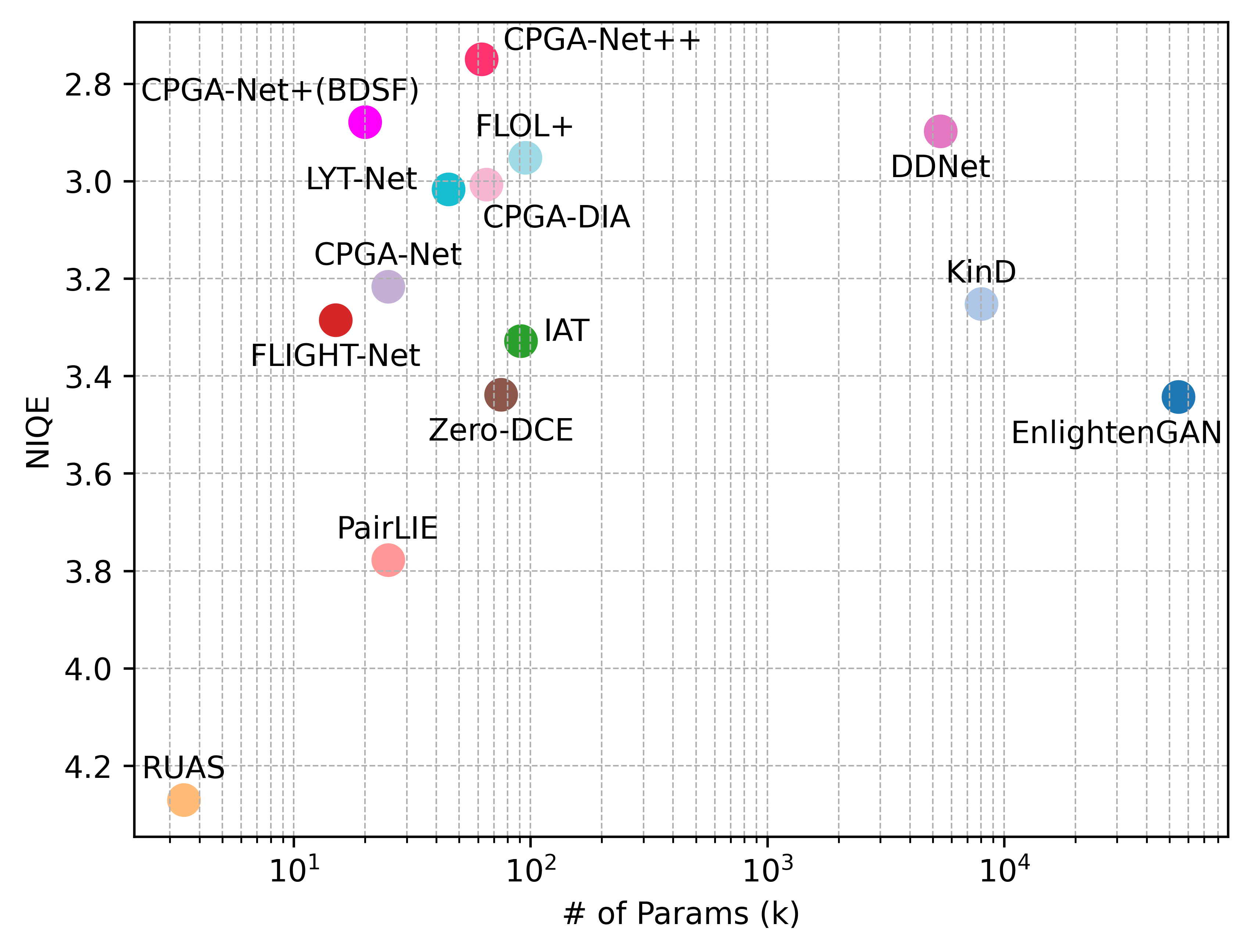}
        \caption{}
     \end{subfigure}
    \caption{Performance Comparison with SOTA approaches. (a) Comparison of PSNR vs. Number of Parameters on LOLv1; (b) Comparison of mean NIQE vs. Number of Parameters on unpaired data.}
    \label{fig.Performance_comparison}
\end{figure}


In our in-depth analysis of our previous work, CPGA-Net~\cite{weng2024lightweight}, we observed a significant discrepancy between our initial hypothesis and its empirical behavior. Specifically, the local branch, intended to provide supportive features for the global processing, converged towards a near-identity mapping. This indicates that the local branch offers minimal contribution during inference, rendering it largely redundant in the final design. This critical finding, however, presents two compelling opportunities: leveraging this redundancy for ultra-lightweight structural pruning and unleashing the potential capacity by stimulating it for more powerful performance.

Acting on these insights, this paper introduces a significant evolution of the CPGA-Net framework by addressing these opportunities. We expanded the theoretical structure by transforming the theoretical equations into an attention mechanism, and we have conducted a thorough analysis of the underlying learning mechanisms and their potential. Our main contributions are as follows:

\begin{itemize}
    \item \textbf{Extended CPGA-Net:} We propose an enhanced version of the Channel-Prior and Gamma-Estimation Neural Network (CPGA-Net) called CPGA-Net+. This model achieves state-of-the-art (SOTA) image quality and efficiency performance for supervised and unsupervised learning. It is a lightweight and practical solutions for real-world applications, as shown in Fig.~\ref{fig.Performance_comparison}.
    \item \textbf{Theoretically-based Attention for Illumination:} We modularized the Atmospheric Scattering Model into a block design and incorporated gamma correction into the local branch. This architecture modification significantly improves image structure and detail, maximizing the efficiency of prior knowledge for illumination to strengthen the overall image quality.
    \item \textbf{Insight-Driven Network Simplification and Refinement:} We identify a paradoxical training dynamic within our baseline model, which allows us to propose both a novel, training-free pruning method for simplification and an advanced fusion module that intelligently re-balances the network for enhanced performance.
\end{itemize}

\section{Related Work}
Our work enhances the original purely convolutional architecture by integrating an attention mechanism while preserving the benefits of a lightweight and efficient design. This advancement is particularly well-suited for LLIE tasks. This section will conduct a comprehensive literature review on leveraging a deep learning-based approach in LLIE and explore developments in lightweight model architectures.

\subsection{Deep Learning-Based LLIE}
With the continuous development of LLIE, Retinex theory has increasingly demonstrated its potential in conjunction with deep learning techniques. Several methods based on this approach address low-light environments. For instance, while LIME~\cite{Guo2017lime} differs from directly decomposing images according to Retinex theory, it primarily relies on estimating the illumination map of low-light images for enhancement. RetinexNet~\cite{Chen2018Retinex} and KinD~\cite{zhang2019kind} decompose images into reflectance and illumination components during decomposition. In the adjustment phase, they adjust the illumination component's brightness and denoise the reflectance component, ultimately merging them based on the theory to restore natural images. EnlightenGAN~\cite{Jiang2021EnlightenGAN} proposes an unsupervised Generative Adversarial Network (GAN) that can be trained without paired low/normal light images. It introduces a global-local discriminator structure, self-regularized perceptual loss fusion, and attention mechanisms to enhance image quality. LLFlow~\cite{wang2021llflow} presents a flow-based framework for LLIE that uses an illumination-invariant color map instead of a Gaussian distribution. It features an encoder for stable color attribute extraction and an invertible neural network to transform low-light images into normally exposed distributions, improving image quality and visibility. LLFormer~\cite{wang2023llformer} integrates a specialized illumination attention module and adaptive exposure fusion, allowing it to dynamically adjust to varying lighting conditions. While these methods effectively addressed the LLIE problems, they overlooked crucial computational issues essential for practical implementation.

\subsection{Lightweight LLIE}
In the context of LLIE, developing lightweight methods is crucial for practical deployment, often requiring sophisticated techniques to achieve both efficiency and effectiveness. For example, Zero-DCE~\cite{guo2020zerodce} replaces the direct image enhancement process with a curve-fitting approach, introducing a series of reference-free loss functions that reduce the computational burden, achieving an efficient and lightweight design. RUAS~\cite{liu2021ruas} builds upon Retinex theory by proposing a Retinex-inspired model that leverages prior information from low-light images, combined with a distillation unit-based search architecture and a cooperative bilevel search strategy, maintaining high performance while achieving a lightweight design. IAT~\cite{Cui_2022_IAT} decomposes the task into local and global processing components. The local branch leverages a convolution-based Transformer to perform image restoration and enhancement. In contrast, the global branch utilizes global priors, including color transformation matrices and gamma correction, to apply global adjustments across different exposure conditions, thereby attaining efficient and lightweight performance improvements. PairLIE~\cite{fu2023pairlie} deviates from the traditional Retinex approach of directly decomposing images; instead, it removes noise through a self-supervised mechanism before decomposition. It shows that training on low-light images of the same scene with different exposures learns features better. Finally, it merges them using a simple convolutional network to achieve a lightweight design. Inspired by Retinex theory and ISP (Image Signal Processor) frameworks, FLIGHT-Net~\cite{ozcan2023flight} features Scene Dependent Illumination Adjustment for illumination and gain processing and Global ISP Network Block for compact color correction and denoising. This design optimizes for both efficiency and lightweight operation. LYT-Net~\cite{brateanu2025lyt} is a recent advancement in LLIE that enhances efficiency by operating in the YUV color space. It uses a lightweight convolutional module to process the luminance (Y) channel with a Transformer block while denoising the chrominance channels (U/V). This approach significantly reduces computational costs and illustrates how thoughtful design integrates Transformer modeling with real-time performance.

\subsection{Insights and Innovations}
Our method is inspired by two foundational insights from prior literature: the effectiveness of theory-guided attention mechanisms and the importance of lightweight design for practical LLIE. Drawing from principles observed in Retinex-based approaches~\cite{Guo2017lime, Chen2018Retinex, zhang2019kind} and atmospheric scattering models (ATSM) ~\cite{Dong2011Fast, weng2024lightweight, weng2024dia}, we reinterpret the interaction between light and image degradation—particularly the way visual information is distorted by ambient air turbulence or illumination imbalance—as a guiding prior for attention learning. This perspective informs the design of our attention module, which selectively emphasizes structurally and visually significant regions during enhancement. Unlike conventional attention mechanisms that rely solely on data-driven cues, our model incorporates domain-specific priors ~\cite{Shi2018Nighttime, weng2024lightweight, brateanu2025lyt} to guide feature modulation. This allows for better contrast enhancement and preservation of fine-grained details, even under challenging lighting conditions. Additionally, building upon the practices in \cite{weng2024lightweight, Cui_2022_IAT, ozcan2023flight}, we integrate gamma correction not just as a global post-processing step, but as a learnable and spatially adaptive component within our network. This shift from global to localized gamma adjustment increases the model’s adaptability to varying exposure and illumination distributions across an image.

Overall, our architecture exemplifies the fusion of physical imaging principles with deep learning efficiency. By embedding theoretical insights into a compact and modular attention framework, our model achieves high-quality enhancement with minimal computational overhead—making it suitable for real-world low-light applications on edge devices.

\section{Methodology}
\begin{figure*}[th]
    \centering
    \includegraphics[width=\linewidth]{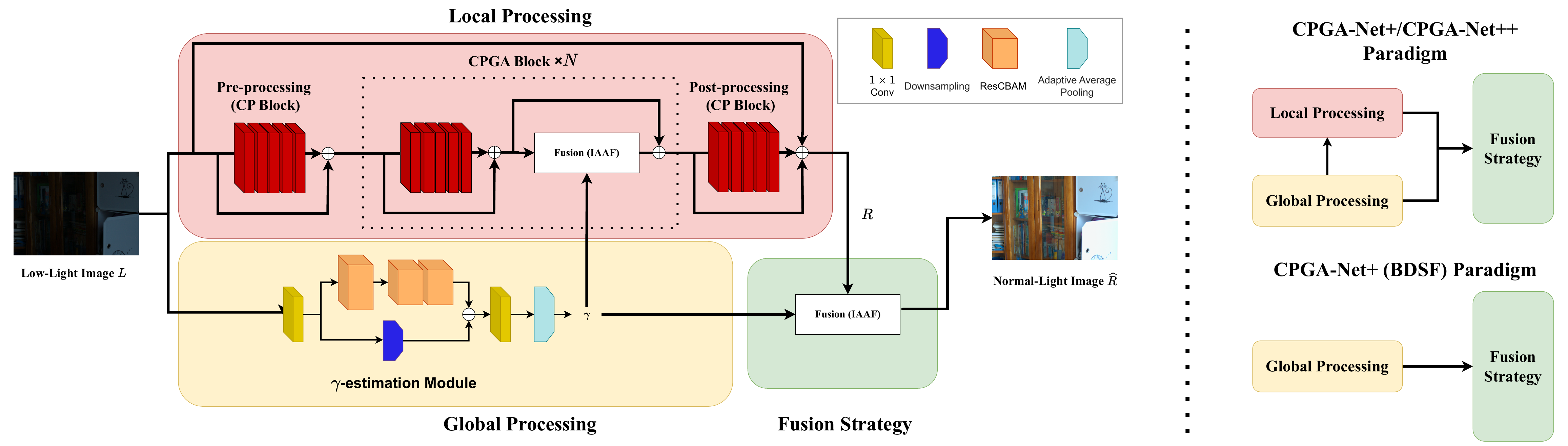}
    \caption{The schematic of our proposed approach, CPGA-Net+. This paper presents the baseline model CPGA-Net+ and two improvements: CPGA-Net+ (BDSF) for an ultra-lightweight design and CPGA-Net++ for enhanced performance, as detailed on the right.}
    \label{fig.cpganet_p}
\end{figure*}

In this section, we delve into the reconstruction of ideal images by global and local concepts in image processing, leveraging advanced deep learning techniques. The discussion will commence with the theoretical underpinnings and motivations for developing CPGA-Net+, followed by an exposition of the network's architecture and implementation, which can be separated as Atmospheric Scattering-driven Attention and Plug-in Attention with Gamma Correction, as shown in Fig.~\ref{fig.cpganet_p}.

\subsection{Preliminary}
Our research is based on a key finding by Guo et al.~\cite{Guo2017lime}, which indicates that a strong theoretical relationship exists between Retinex theory \cite{land1977retinex} and the Atmospheric Scattering Model (ATSM) \cite{mccartney_optics_1976}. This connection allows for a reinterpretation of LLIE through the lens of haze removal principles. Retinex theory assumes that the received image can be decomposed into illumination and reflectance components, as shown in Eq. (1):

\begin{equation}
    S = i \cdot R,
    \label{eq.retinex}
\end{equation}
where \(S\) is the perceived image, \(i\) denotes the illumination component, and \(R\) represents the reflectance component.

On the other hand, the Atmospheric Scattering Model for haze removal is defined as:

\begin{equation}
    I = t J + (1-t) A
    \label{eq.ATSM}
\end{equation}
where \( I \) represents the input image, \( J \) represents the haze-free image, \( t \) represents the atmospheric transmission, and \( A \) represents the intensity of atmospheric light.

Based on Dong et al.~\cite{Dong2011Fast}, the low-light image \(I\) can be seen as \(1 - L\), where \(L\) represents the low-light image, and \(J\) can be seen as \(1 - R\), where \(R\) reflects the important characteristics of the input image. The above substitutions are performed under the normalized pixel values in [0, 1]. With these substitutions, we rewrite Eq.~\eqref{eq.ATSM} into the following form:
\begin{equation}
    R = \tilde{t} L + (1-\tilde{t}) \tilde{A}
    \label{eq.enhance_atsm}
\end{equation}
where \(\tilde{A} = 1 - A\) and \(\tilde{t} =1/t\). The model described in Eq.~\eqref{eq.enhance_atsm} is the cornerstone of our neural network design. Part of the information of the reflectance \( R \) comes from the known image \( L \), and part of it comes from an unknown image \( \tilde{A} \), and the proportion sum of their contributions is limited to 1. When \( L \) is very dark or noisy (the scene information is less reliable), the contribution of \( L \) is lowered, and the contribution of \( \tilde{A} \) is increased; when \( L \) is relatively bright and less noisy (the scene information is more reliable), the contribution of \( L \) is increased, and the contribution of \( \tilde{A} \) is reduced. So, \( \tilde{t} \) should reflect the intensity level of \( L \) in some way.

\begin{figure}[!t]
  \centering
  \includegraphics[width=\linewidth]{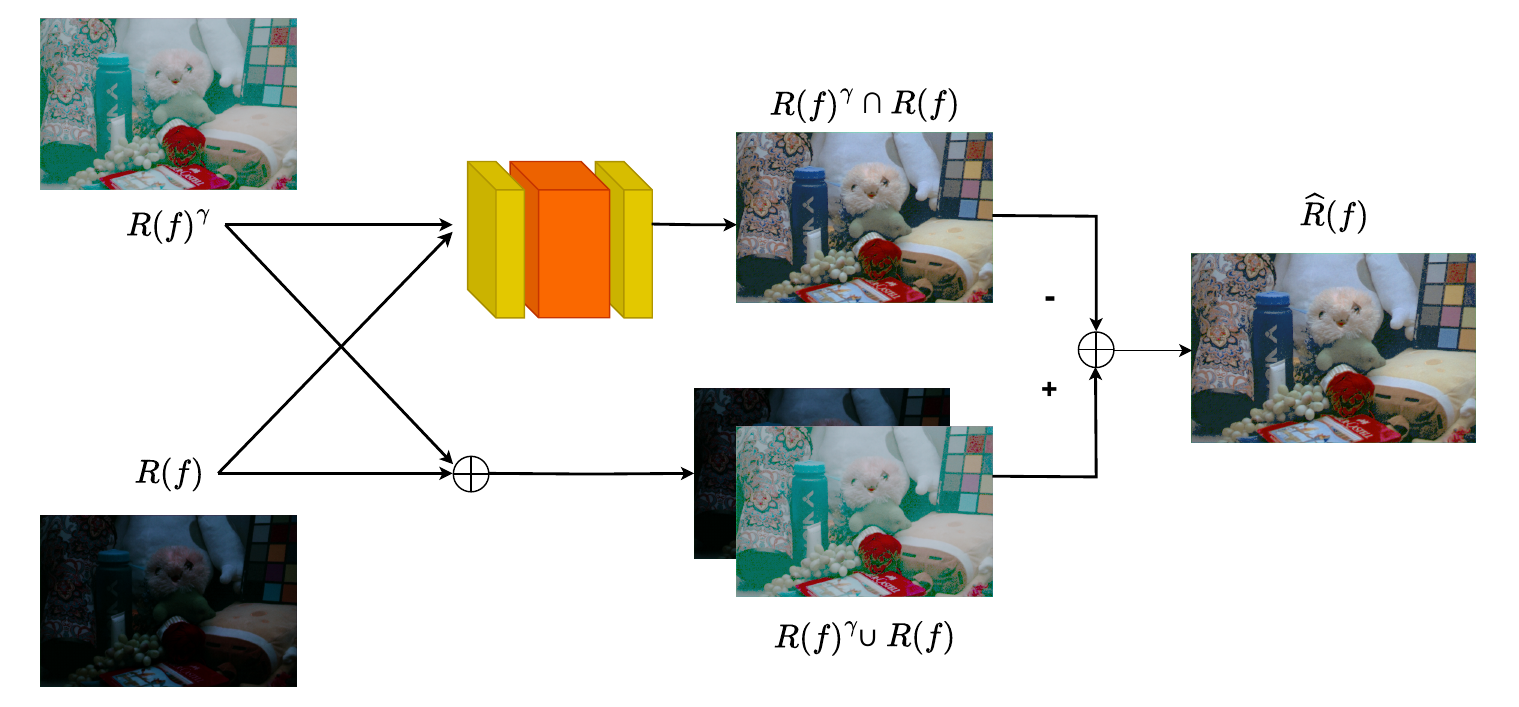}
  \caption{A block diagram of the IAAF (Intersection-Aware Adaptive Fusion) module~\cite{weng2024lightweight}.}
  \label{fig:IAAF}
\end{figure}

In our previous work, CPGA-Net~\cite{weng2024lightweight}, we successfully utilized these theoretical equations in deep learning form as the local processing in image enhancement. Moreover, to further refine LLIE, we also considered global processing technique, gamma correction, as the brightness control, which is a simple technique that adjusts all the pixels with pointwise exponential operations, as shown in Eq.~\eqref{eq.gamma_correction}:

\begin{equation}
    s = r^\gamma,
    \label{eq.gamma_correction}
\end{equation}
where \( \gamma \) is the gamma value controlling the degree of correction, enhancing the input image \( r \) to produce the output image \( s \). These approaches combined an independent branch of regression with the enhancement model to better estimate gamma values. Moreover, the complexity of gamma value estimation can complicate training goals and make the process prone to divergence. Taking these aspects into account, we proposed an IAAF (Intersection-Aware Adaptive Fusion) module, as shown in Fig.~\ref{fig:IAAF} and Eq.~\eqref{eq.IAAF}:

\begin{equation}
    \begin{split}
        \hat{R} &= \text{IAAF}(R^\gamma, R) = (R \cup R^\gamma) - (R \cap R^\gamma) \\ 
                &\approx R + R^\gamma - \cap(R, R^\gamma),        
    \end{split}    
    \label{eq.IAAF}
\end{equation}
where the enhanced image \( \hat{R} \) is created by combining \( R \) and \( R^\gamma \) while removing any overlapping elements and \( \cap(R, R^\gamma) \) represents the intersection estimation for finding similar features across gamma-corrected and uncorrected images. CPGA-Net enhances image quality and contrast with good perspective evaluation by combining local and global processing.

\subsection{The Connection Between Theoretical Equations and Low-Light Image Enhancement}
In this section, we thoroughly evaluate the CPGA-Net from the block structure perspective, including reevaluating the application of prior knowledge as Atmospheric Scattering-driven Attention and the mixture of local-global processing as Plug-in Attention with Gamma Correction. 

\subsubsection{Atmospheric Scattering-driven Attention}
First of all, the reformulation of Eq. (3) reveals an alternative imaging perspective, where L is linked to the characteristic of atmospheric light, which predominantly includes environmental interference, $\tilde{A}$ corresponds to the unknown noise-free image, and R corresponds to the reflectance in a linear relationship between $L$ and $\tilde{A}$. This formulation's underlying mechanism matches the phenomenon of our atmospheric scattering-based approach, performing an insightful attention relationship of these factors. Further analysis will be discussed in the Appendix.

Built upon these relationships, we extended the local processing as an attention mechanism called the Channel-Prior block (CP block), which modularizes the relationship of underlying equations into a systematic form. We restructured the module with convolutions and a ResBlock~\cite{he2016resnet}, fusing features with the original RGB channels to streamline it into an attention-block design. 

For the attention to the brightness correlated priors for transmittance $\tilde{t}$ estimation, we adopt channel-prior features $F^{CP}$, which are sensitive to contrast and brightness variations. These are derived from the input feature map $f$:

\begin{equation}
    F^{\text{CP}}(f) = \text{concat} \left( \max_c \left(f^c\right), \min_c \left(f^c\right), \underset{c}{\text{mean}} \left(f^c\right) \right),
\end{equation}
where \(f\) is the input feature map, and \(F^{\text{CP}}\) denotes the channel prior features, which include brightness sensitive channel priors: Bright Channel Prior (BCP), Dark Channel Prior (DCP), and the luminance channel (I component in the HSI color space). These priors have been widely used in prior research to capture the characteristics of low-light images. In our implementation, we simplify the prior into a high-dimensional representation by reducing the channel priors to the input channels' maximum, minimum, and mean values. This approach improves the attention module's responsiveness in controlling the image's overall contrast.

For the \(\tilde{A}\) estimation, which captures detailed features and reconstructs the image, we redesigned it as a mini-U-Net-based architecture with encoder and decoder pathways connected by skip connections. This approach ensures that the model remains efficient and suitable for real-time applications or scenarios with limited computational resources without significantly compromising the quality of the reconstructed image.

\begin{figure}[!t]
  \centering
  \includegraphics[width=\linewidth]{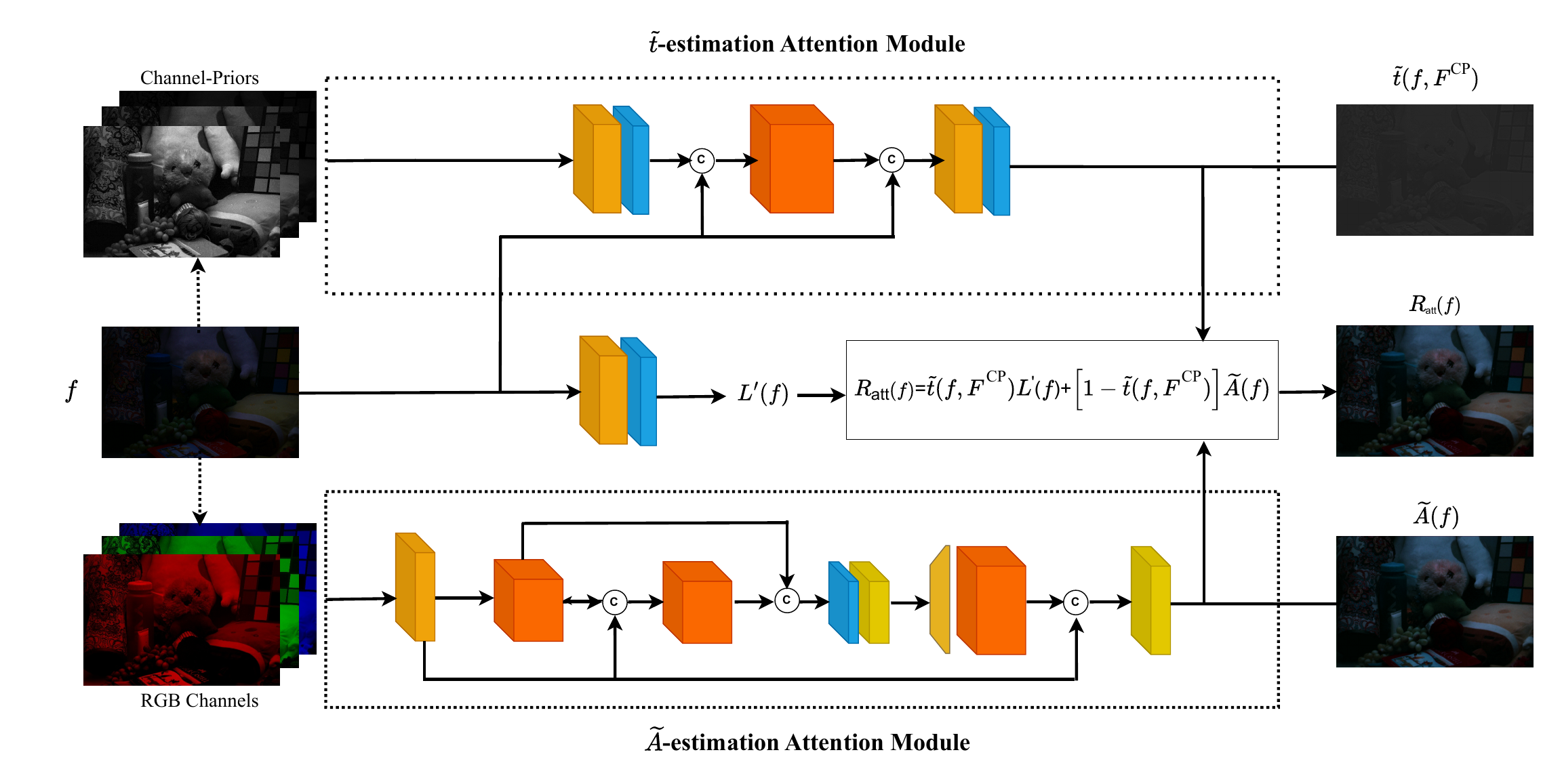}
  \caption{The Channel-Prior Block (CP block) block diagram, where c denotes a concatenation operation.}
  \label{fig:CP_block}
\end{figure}

After obtaining the estimates of \(\tilde{t}\) and \(\tilde{A}\), we can reconstruct our features using Eq.~\eqref{eq.enhance_atsm}, which serves as an attention module sensitive to brightness variations in the scene. This leads to the proposed Atmospheric Scattering-driven Attention, formulated as follows:
\begin{equation}
    R_{\text{att}}(f) = \tilde{t}(f, F^{\text{CP}}) L'(f) + \big[1 - \tilde{t}(f, F^{\text{CP}})\big] \tilde{A}(f),
    \label{eq.ASD attention}
\end{equation}
where \(L^{'}\) is the mapped input tensor with a matching channel for formula calculation, $\tilde{t}(f,F^\text{CP})$ indicates the derived transmittance with input feature map $f$ and channel-prior features $F^\text{CP}$, and the entire structure is shown in Fig.~\ref{fig:CP_block}.

\subsubsection{Plug-in Attention with Gamma Correction}
As analyzed in \cite{weng2024dia}, the adaptive gamma value serves as an environmental factor representing the overall illumination conditions, directly reflecting the restoration's state. To leverage the prior information of environment representation, we have integrated a global processing operation within the Channel-Prior Block and incorporated it directly into the attention mechanism. This approach enables the global feature to estimate the optimal gamma-correction value for each feature channel, rather than limiting gamma correction to just the final step, as is often done in previous studies \cite{weng2024lightweight, Cui_2022_IAT}. The modified attention can then be expressed as:

\begin{equation}
    \hat{R}_{\text{att}} = \text{IAAF}(R_{\text{att}}^\gamma, R_{\text{att}}) + R_{\text{att}},
    \label{eq.plugin_attention}
\end{equation}
where \( \hat{R}_{\text{att}} \) is the attention feature built on the gamma corrected feature \( R_{\text{att}}^\gamma \) and the uncorrected feature \( R_{\text{att}} \). Moreover, we add another residual to this attention in Eq.~\eqref{eq.plugin_attention}, distinguishing it from the reconstruction applied in Eq.~\eqref{eq.IAAF}, making it an auxiliary attention to support the residual features.

Consequently, this attention module ensures that the network focuses on regions within the broader image where gamma correction yields the most significant enhancement. This approach is called ``Plug-in Attention with Gamma Correction,'' and the CP block with IAAF is named  ``CPGA block.''

\subsection{Exploiting Implicit Regularization of CPGA-Net}
While our prior work, CPGA-Net~\cite{weng2024lightweight}, demonstrated strong performance by adhering to a theoretical image restoration model, a deeper analysis of its internal mechanisms reveals a critical limitation. As illustrated in Fig.~\ref{fig:IAAF}, the feature maps generated by the local branch are nearly identical to the input, indicating that the network learns to approximate a near-identity mapping in this path. This situation creates redundancy and suggests the system fails to provide localized feature corrections. The main reason came from the design of the fusion strategy.

\begin{figure}
     \centering
     \begin{subfigure}[b]{0.45\linewidth}
         \centering
         \includegraphics[width=\textwidth]{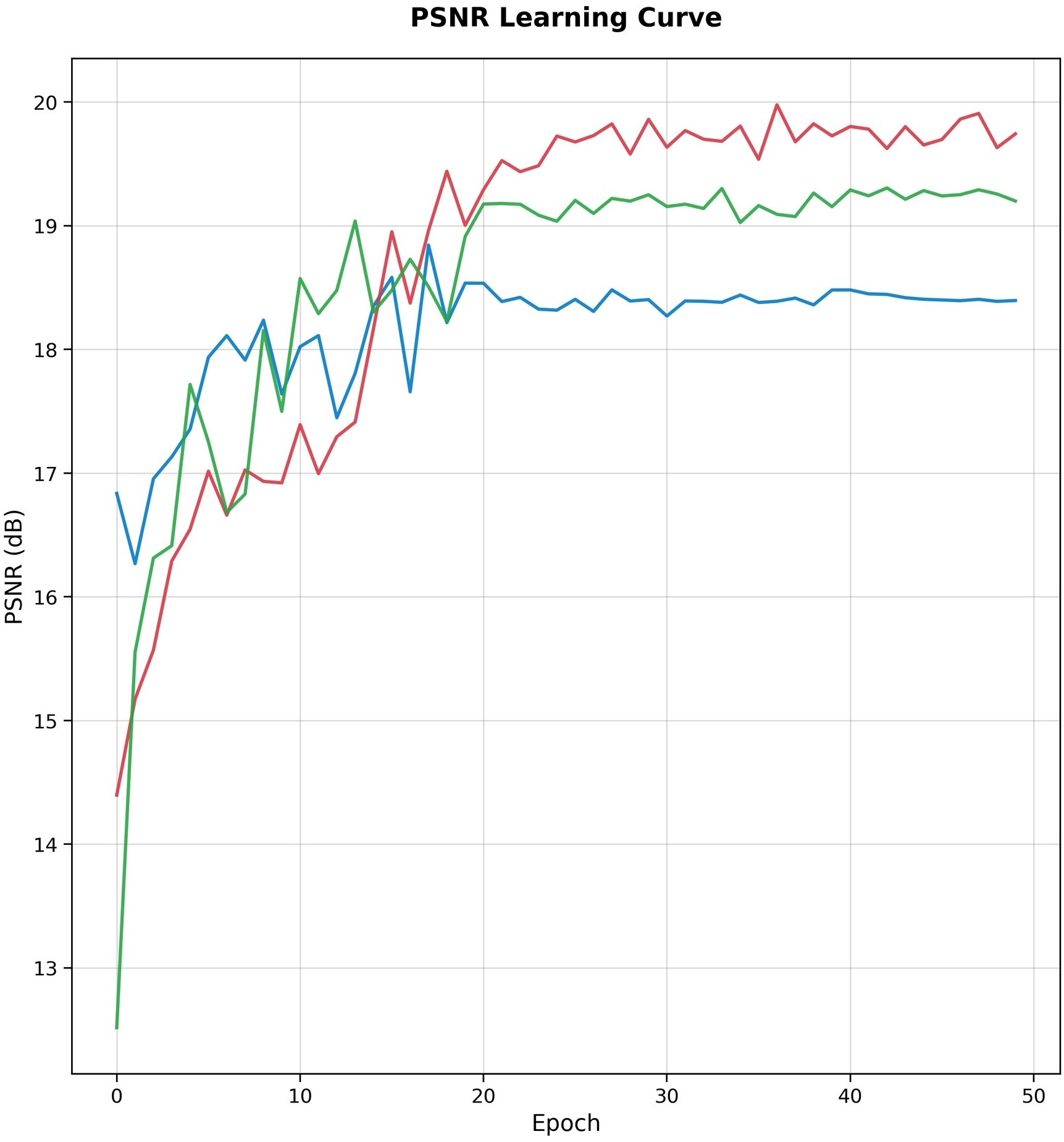}
         \caption{PSNR}
     \end{subfigure}
     \begin{subfigure}[b]{0.45\linewidth}
         \centering
         \includegraphics[width=\textwidth]{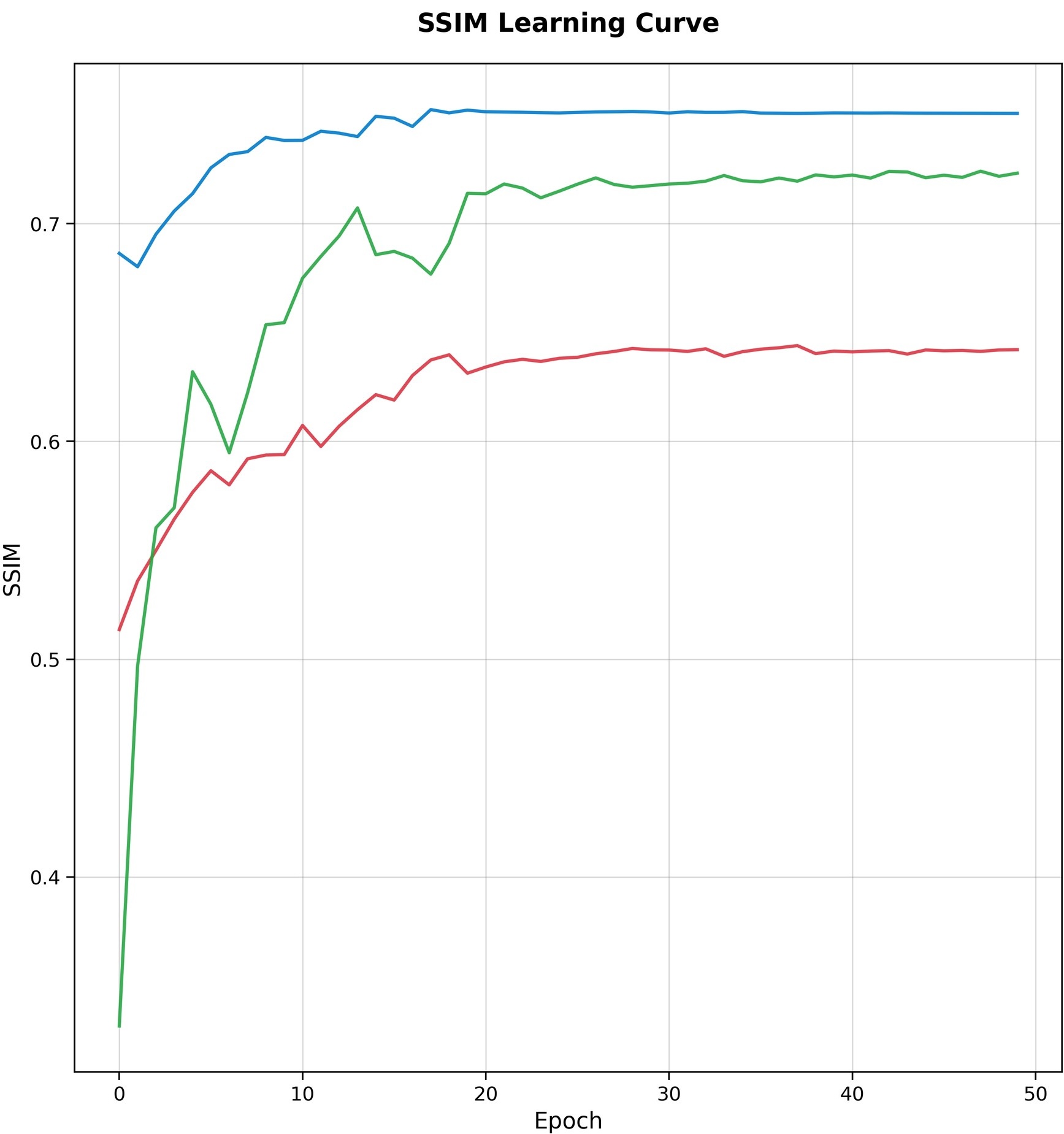}
         \caption{SSIM}
     \end{subfigure}
     \hfill
    \caption{The learning curve comparisons of ablation results for CPGA-Net~\cite{weng2024lightweight}. 
    \textcolor{blue}{Blue} represents the local branch only, 
    \textcolor{red}{red} represents the global branch only, and 
    \textcolor{green!50!black}{green} represents the complete CPGA-Net.}
    \label{fig:lr_cpganet}
\end{figure}

IAAF is an effective fusion strategy inspired by set theory, designed to blend two feature sets by compensating for their union with a learned intersection. Our primary objective was to employ IAAF to integrate the base reflectance \( R \) and its gamma-corrected counterpart \( R^\gamma \). The global branch manages brightness control, and the local branch concentrates on detail reconstruction. Our analysis in Fig.~\ref{fig:lr_cpganet} revealed a significant issue with asynchronous convergence between the two branches. The global branch handles brightness adjustment through factorized gamma correction, which enhances contrast but often fails to capture fine details. In contrast, the local branch emphasizes restoring structural elements and delivers superior performance, although it presents greater training challenges.

\begin{figure}[t]
    \centering
    \includegraphics[width=\linewidth]{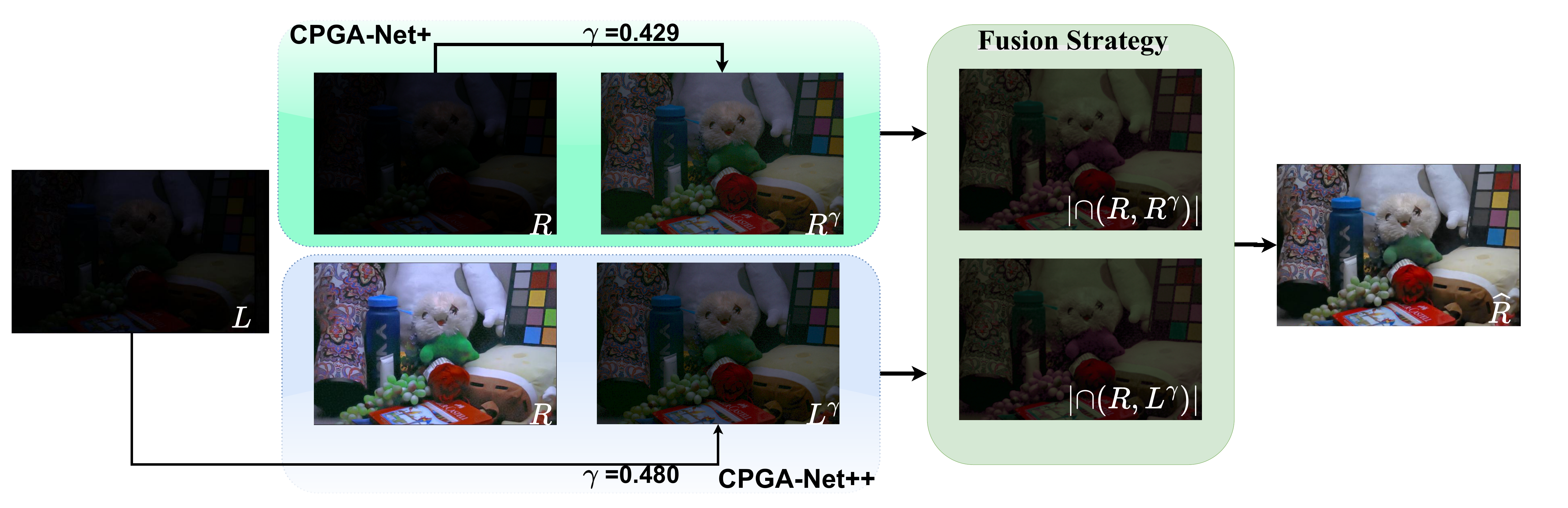}
    \caption{Visual analysis of the intermediate feature maps and fusion strategies in CPGA-Net+ and CPGA-Net++.}
    \label{fig:VA_cpga}
\end{figure}

While merging branches can yield a balanced quality, this discrepancy creates a training imbalance. Consequently, the final output of the local branch tends to be a suboptimal compromise, with global processing dominating the restoration process, leading to a critical challenge, as the dominance of the global branch suppresses the local branch, stabilizing the restoration but ultimately impeding overall performance. This suppression results in a near-identity mapping of the local branch, even while achieving improved outcomes, as shown in Fig.~\ref{fig:VA_cpga}.

This analysis uncovers a fascinating and counterintuitive dynamic: \textbf{an adversarial yet ultimately symbiotic relationship between the global and local branches}. While converging to a near-identity mapping, the local branch introduces a subtle disturbance to the feature space. Crucially, instead of being a flaw, this disturbance acts as a powerful form of implicit regularization, supporting the global branch by forcing it to learn a more robust and generalized restoration mapping. Consequently, the entire restoration workload becomes centralized within the global branch, which, strengthened by this adversarial process, achieves a more stable and effective final solution. 
This regularization dynamic is crucial during training, but once the model is trained, the local branch's function reverts to a simple, computationally redundant identity mapping during inference. 

Therefore, we proposed a structural pruning technique based on the concept of block pruning by directly removing the local branch while inferencing because of this characteristic, but it differs from existing methods~\cite{wu2023block, chen2018shallowing, elkerdawy2020filter, sun2024multi} and maintains its simplicity without requiring any additional training. We named this technique ``Block Design Simplification (BDSF).''

\subsection{Unleashing the Power of Intersection-Aware Adaptive Fusion}
In contrast to the pruning, we further explore the solution to address the mismatch of data flow to utilize the overall capacity hidden in the local branch. We introduce IAAF+, a novel module designed to control the fusion process intelligently. We actively apply gamma correction to the input $L$ to address training imbalances, allowing for independent gradient flow. Then, the module operates through two core mechanisms: Weighted Feature Selection (WFS), which models the intersection of features, and Intersection Representation Scoring (IRS), which supplies adaptive fusion weights.

The Weighted Features Selection (WFS) module acts as a dynamic scoring function. It analyzes features from both branches to determine their optimal blending ratio, outputting a single learnable scalar parameter $\alpha$. This value is then used to compute a baseline feature union ${\hat{R}}_{\cup}$, as a weighted average of the two branches:
\begin{equation}
    \hat{R}_\cup = \alpha R + (1-\alpha)L^\gamma.
\end{equation}
By making the fusion weight $\alpha$ learnable, the WFS module allows the network to balance the contribution of each branch throughout the training process. This creates an adaptive fusion strategy that harmonizes the two branches instead of letting them compete.

Concurrently, the Intersection Representation Scoring (IRS) module explicitly models the shared, redundant information with more accurate and detailed mapping in the intersection of IAAF:

\begin{equation}
    \hat{R}_\cap = \cap(S_R\odot R, S_{L^\gamma}\odot L^\gamma),
\end{equation}
where $S_R$ and $S_{L^\gamma}$ are the spatial attention maps produced via $R$ and $L$, respectively. Each feature map is modulated via element-wise multiplication $\odot$ with the corresponding spatial attention maps, allowing the intersection to emphasize reliable regions while suppressing noise, resulting a cleaner and more informative shared representation of fusion. This allows the module to perform a selective and weighted intersection calculation, focusing only on each branch's most reliable feature regions to define the shared information. This prevents noisy or less relevant features from corrupting the intersection estimate, leading to a more precise and meaningful compensation in the final fusion step. 

\begin{figure}
    \centering
    \includegraphics[width=\linewidth]{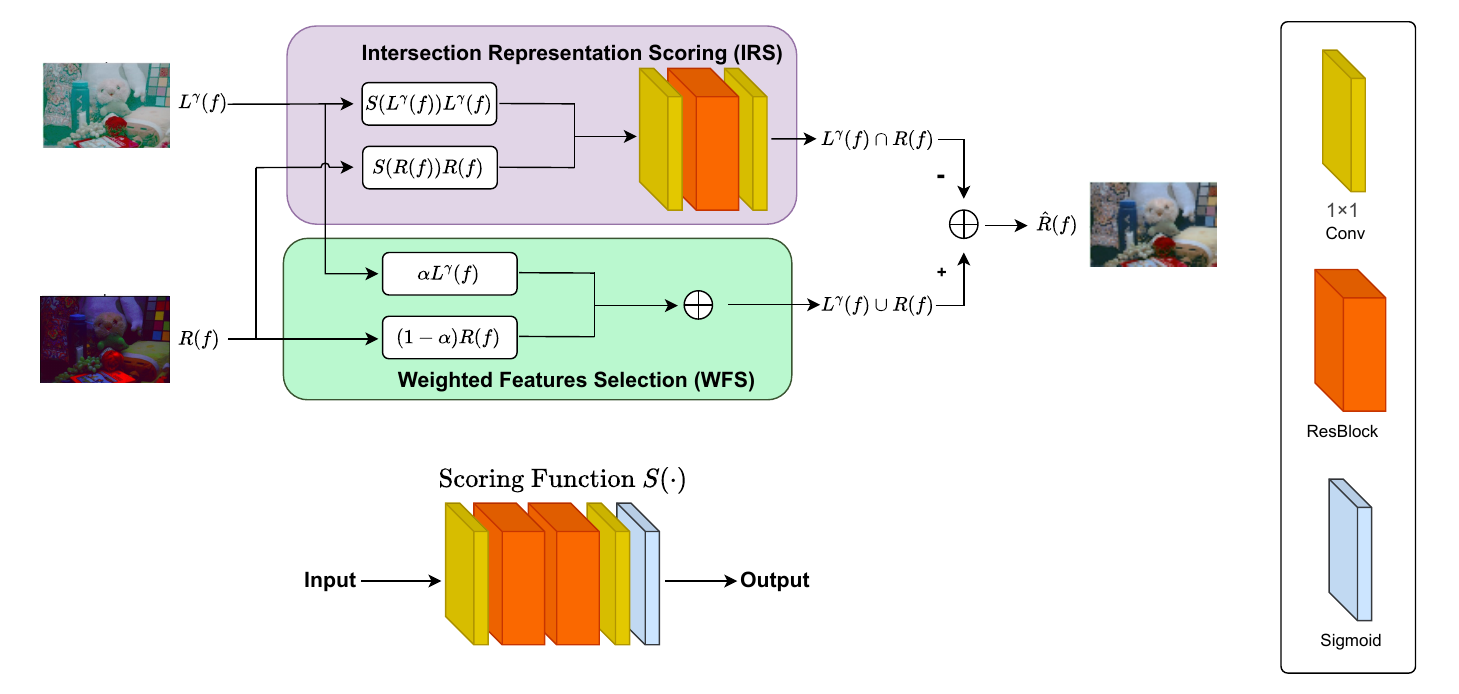}    
    \caption{A block diagram of the IAAF+ module. $\alpha$ is the learnable scalar parameter to identify the importance of input features toward fusion by the WFS module. In contrast, the IRS module uses the spatial attention maps $S$ to evaluate the effectiveness of input signals for intersection representation.}
    \label{fig:iaaf_masking}
\end{figure}

Overall, the final output, $\hat{R}(x)$, is produced by first creating a weighted average of the two branches controlled by WFS, and then subtracting the intersection calculated by IFS. This entire process is captured in the following formulation:
\begin{equation}
    \hat{R}_{\text{IAAF+}} = \hat{R}_\cup - \hat{R}_\cap ,
\end{equation}
which separates the task of global balancing (via WFS's $\alpha$) from specific feature compensation (via IRS's intersection map), promotes a synergistic relationship between the branches for a more robust and detailed final image. The diagram of IAAF+ is shown in Fig.~\ref{fig:iaaf_masking}.

Our module has a two-phase architecture that creates a deliberate separation of concerns. The intermediate stages focus on self-enhancement, where the network iteratively refines the global feature representation $R$ by learning to adjust its illumination and contrast against its gamma-corrected version. The final stage then performs the critical heterogeneous fusion, injecting the fine-grained details from the separate local branch $L^\gamma$ only after the global feature $R$ has been fully stabilized and optimized. The overall analysis compared to baseline is shown in Fig.~\ref{fig:VA_cpga}.

\subsection{Loss Functions}
In this section, we present the training methodology of our approach, which incorporates both supervised and unsupervised learning strategies. The unsupervised scheme is designed to induce pseudo-supervised behavior, thereby extending the applicability of our method to real-world scenarios without ground-truth supervision.

\subsubsection{Supervised learning}
In supervised learning, we use four loss functions to guide our approach: L1 loss, perceptual loss, SSIM loss, and HDR L1 loss. 

The L1 loss function, a commonly used loss function that performs better in image enhancement and restoration, is defined as:
\begin{equation}
    L_1 = \| \hat{Y} - Y^\text{GT} \|_1
\end{equation}
where \(\hat{Y}\) is the output and \(Y^\text{GT}\) is the ground truth.

Perceptual loss~\cite{johnson2016perceptual} is commonly used in image restoration, style transfer, and generation. It emphasizes capturing high-level features and structures that closely resemble human perception. The loss can be expressed as:

\begin{equation}
    L_{per} = \| \Psi(\hat{Y}) - \Psi(Y^\text{GT}) \|_2^2
\end{equation}
where \(\Psi\) represents the feature extractor of VGG16.

SSIM loss is a function that measures the similarity between two images based on structural information via the SSIM index (structural similarity index). It compares luminance, contrast, and structure, reflecting perceptual quality better than traditional pixel-wise losses. It can be written as:
\begin{equation}
    L_{\text{SSIM}} = 1 - \text{SSIM}(\hat{Y}, Y^\text{GT})
\end{equation}

To better handle HDR image content, we further adopt the HDR L1 loss. As introduced in \cite{liu2022ghost}, it is computed in the tone-mapped domain since HDR (High Dynamic Range) images are typically viewed after tone-mapping. To achieve this, they apply the widely used µ-law function to calculate the loss:
\begin{equation}
    T = \text{sgn} \frac{\log(1+\mu x)}{\log(1+\mu)}
\end{equation}
where \(\mu\) is set to 5000, \(T\) is the tone-mapped HDR image, and \(x\) is the input image. Then, \(\mu\)-law function, we utilized the L1 loss as follows:
\begin{equation}
    L_{\text{HDR-L1}} = \| T(\hat{Y}) - T(Y^\text{GT}) \|_1
\end{equation}
It represents the image in the tone-mapped domain, ensuring the loss is calculated in a perceptually relevant space that aligns with how normal light images are typically viewed.

\subsubsection{Unsupervised training strategy}
While our supervised learning approach yields effective image enhancement, a fully supervised method becomes infeasible when ground truth normal-light images are unavailable. To address this limitation, we propose an unsupervised training strategy aligned with our overall network framework, which effectively balances contrast, perceptual quality, and smoothness, offering a robust alternative when ground-truth supervision is unavailable. For our unsupervised learning process, we utilize HDR L1 loss, perceptual loss, and total variation loss, with the target being the histogram-equalized low-light image.

First, we use histogram-equalized images with HDR L1 and perceptual loss to adjust tone mapping and content. This guides the network toward producing outputs with appropriate tone mapping. Next, inspired by findings in \cite{weng2024lightweight}, we supervise mid-feature maps using a blurred low-light input with HDR L1 loss. Thus, we utilized total variation loss~\cite{RUDIN1992tvloss, wang2021ISRsurvey} for denoising and improving the smoothness:
\begin{equation}
L_{TV} = \frac{1}{h w c} \sum_{i,j,k} \sqrt{\left(\hat{Y}_{i,j+1,k} - \hat{Y}_{i,j,k}\right)^2 + \left(\hat{Y}_{i+1,j,k} - \hat{Y}_{i,j,k}\right)^2}
\end{equation}
where $h$, $w$, and $c$ represent the height, width, and number of channels, respectively; $i$, $j$, and $k$ represent the indices corresponding to height, width, and channel, respectively.

This unsupervised learning strategy enables our model to generalize to real-world low-light conditions by pseudo-supervised behavior without access to reference images.

\section{Experiment Results}
This section compares our approach with several SOTA methods on benchmark datasets, including paired and unpaired datasets.

\subsection{Datasets and Evaluation Metrics}
For evaluation, we apply our approach to both paired and unpaired datasets. For paired data, we use the LOLv1 and LOLv2 datasets~\cite{Chen2018Retinex}, benchmarks for the LLIE task. LOLv1 includes 485 images for training and testing, while LOLv2 consists of two subsets: real-captured and synthetic. The real-captured subset (LOLv2 Real) has 689 images for training and 100 for testing, while the synthetic subset (LOLv2 Synthetic) has 900 training images and 100 testing images. For unpaired data, we utilize five datasets: LIME~\cite{Guo2017lime}, MEF~\cite{Ma2015MEF}, NPE~\cite{Wang2013NPE}, VV~\cite{Vonikakis2018VV}, and DICM~\cite{Lee2013DICM}. Since these datasets lack ground truth references for paired evaluation, we assess performance using the NIQE metric, which is widely used to evaluate the naturalness of images. For object detection applications, we apply our approach to the Exclusively Dark (ExDark) dataset~\cite{loh2019exdark}, which consists of 7,363 low-light images from very low-light environments to twilight with 12 object classes and provides a suitable benchmark for evaluating object detection performance in such conditions.

\subsection{Implementation Details}
Our experiments are conducted and evaluated with an NVIDIA GeForce RTX 3090 GPU. For the preprocessing of training, we cropped the image into \(256 \times 256\) pixels with random translation. We set 600 epochs for training. An initial learning rate of $10^{-3}$ is set for training with the Adam optimization scheme, and the learning rate changing cycle is 67 epochs by the Cosine Annealing scheduler. To manage computational demands, we introduced a method called ``Block Design Simplification (BDSF)'' in our lightweight version, which directly removes blocks in the local branch during inference. For CPGA-Net++, we utilized IAAF+ to replace IAAF~\cite{weng2024lightweight} and substituted ResBlock~\cite{he2016resnet} with ConvNeXT block~\cite{liu2022convnet}. Due to the ConvNeXT structure, the training of CPGA-Net++ was changed to AdamW with a cosine annealing scheduler until completion. The training of CPGA-Net++ also includes an intermediate supervision of 5 epochs to guide local processing toward the ground truth, ensuring effective learning for the local branch. We have also applied our approach as unsupervised learning using histogram equalization, demonstrating its flexibility. When using unsupervised learning, we trained on LOLv2 Real for 50 epochs, which consists of more realistic data and does not require any normal exposure images for training. 

\subsection{Evaluation Results}
\begin{figure}[t]
    \centering
    \includegraphics[width=\linewidth]{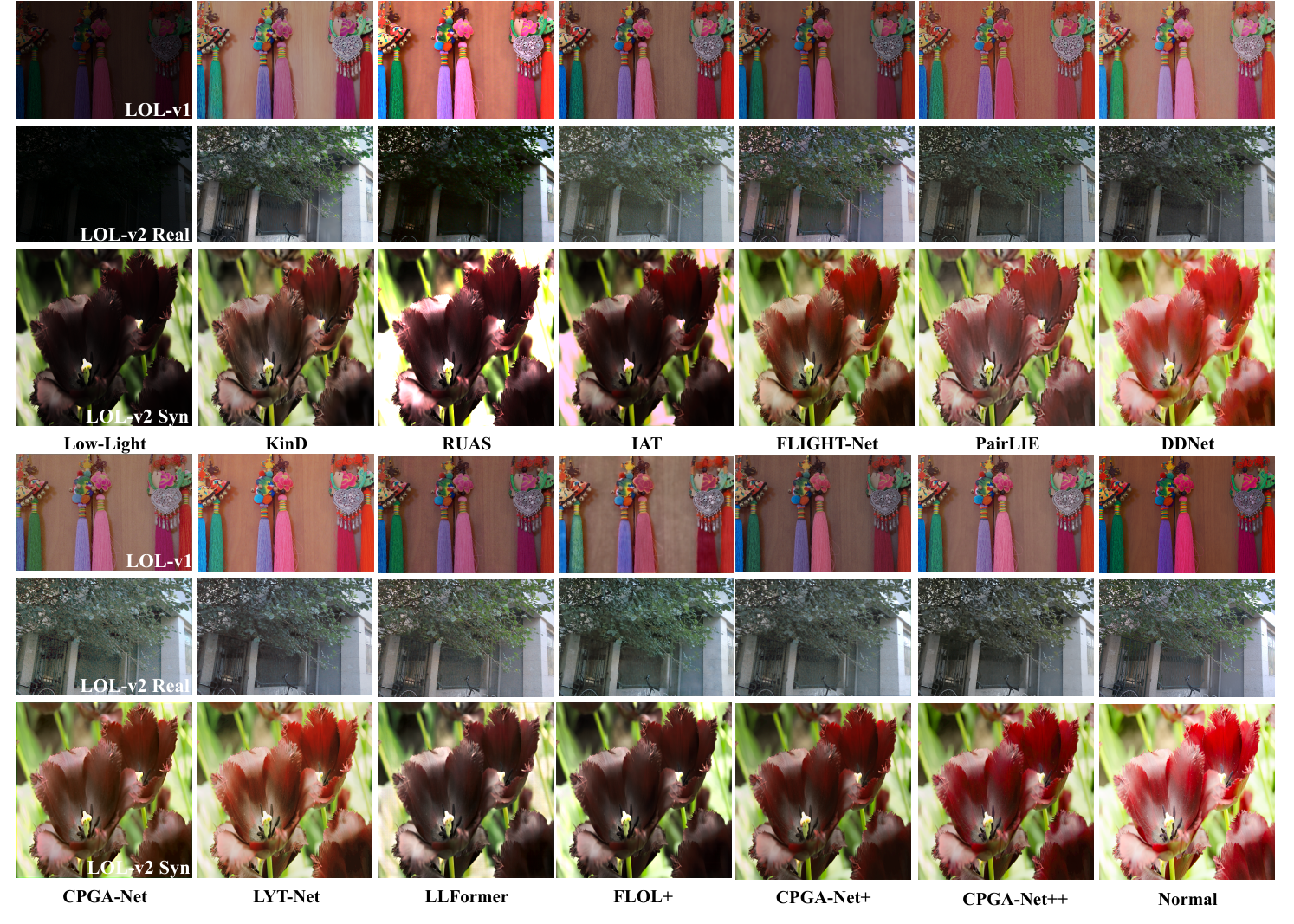}
    \caption{Visual comparison on paired datasets~\cite{Chen2018Retinex}. Zoom in for a better view.}
    \label{fig:lol_fig}
\end{figure}

\begin{figure}[t]
    \centering
    \includegraphics[width=\linewidth]{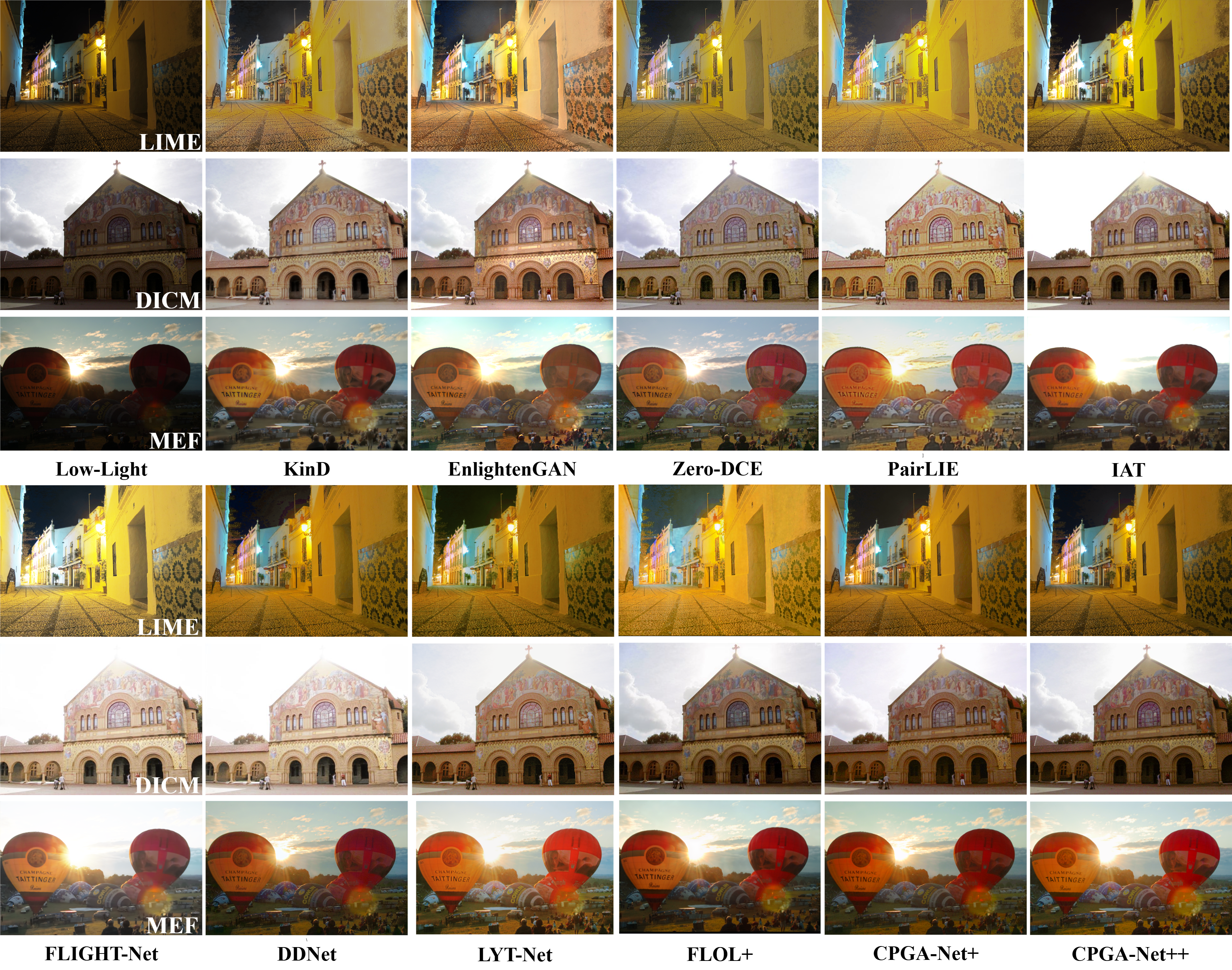}
    \caption{Visual comparison on unpaired datasets~\cite{Guo2017lime, Ma2015MEF, Lee2013DICM}. Zoom in for a better view.}
    \label{fig.niqe_fig}
\end{figure}

Tables~\ref{table.comp_pair} and \ref{table.comp_unpair} show that, our approach achieves a higher standard than other methods, ranking second and third on paired datasets and first on five sets of unpaired data. Furthermore, our BDSF model maintains an ultra-lightweight design with a low number of parameters and FLOPs, resulting in only a minimal decrease in performance. Meanwhile, CPGA-Net++ is significantly enhanced by leveraging the potential of the local branch, thereby achieving strong quality and competitive performance. Visual comparisons are presented in Figs.~\ref{fig:lol_fig} and \ref{fig.niqe_fig}. Also, we figured out that the approaches with CPGA architecture present better quality on unpaired data, which means that our theoretical equations' assumptions for improvement are closer and related to nature, making the image more realistic. Furthermore, our approach improves performance by 0.05 SSIM over CPGA-Net using the same architecture, while maintaining a lightweight design. Notably, CPGA-based methods consistently yield better quality on unpaired data, suggesting that our equation-driven design aligns more closely with natural image properties, resulting in more realistic enhancements. 

\begin{figure}[t]
    \centering
    \includegraphics[width=\linewidth]{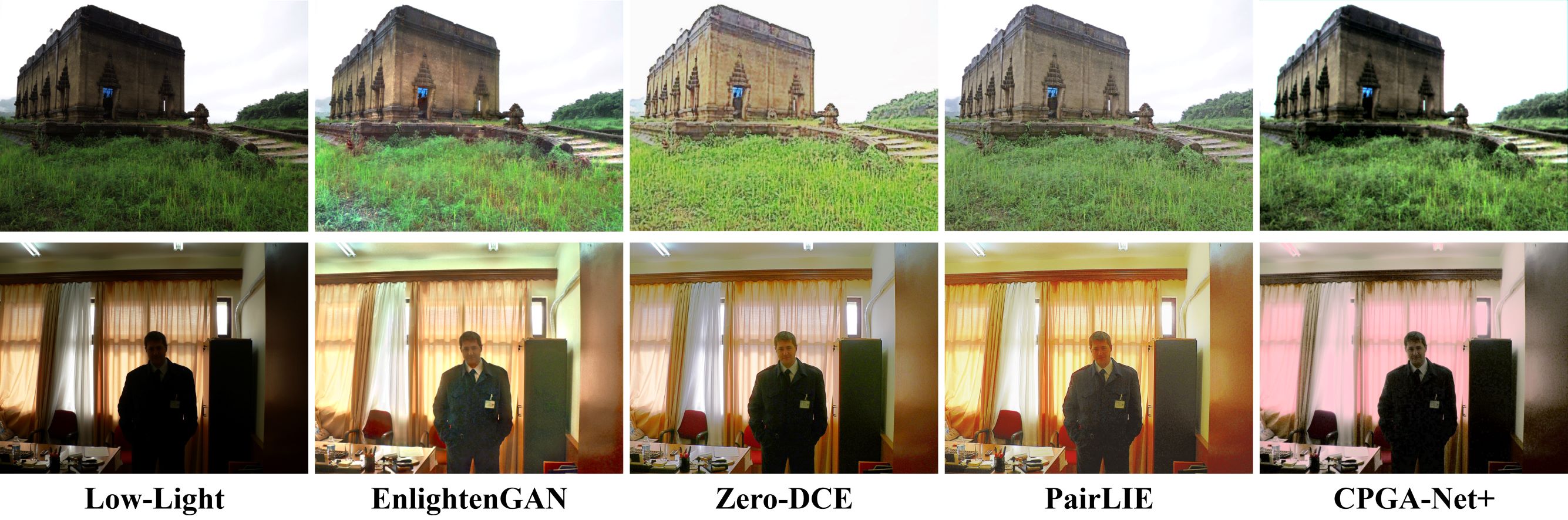}
    \caption{Visual comparison of unsupervised approaches on unpaired data~\cite{Wang2013NPE, Vonikakis2018VV}.}
    \label{fig.unsupervised_fig}
\end{figure}

For the comparison of unsupervised approaches, as shown in Table~\ref{table.comp_unpair}, we ranked first compared to other unsupervised approaches with better contrast. This demonstrates the robustness of our theoretically-based network architecture when using simple supervision of histogram-equalized images. However, there are more noticeable defects and distortions due to the lack of strong supervision of the details,  as shown in Fig.~\ref{fig.unsupervised_fig}. This will be a focus for our future work.

\begin{table*}[!t]
    \centering
    \caption{Comparison to SOTA methods on paired datasets~\cite{Chen2018Retinex}. We represent the first and second ranks with \textbf{bold} and \underline{underlined}, respectively. BDSF means Block Design Simplification for our approach.}
    \label{table.comp_pair}
    \begin{tabular}{lcccccccccc}
        \hline
        \textbf{} & \multicolumn{3}{c}{\textbf{LOLv1}} & \multicolumn{2}{c}{\textbf{LOLv2-real}} & \multicolumn{2}{c}{\textbf{LOLv2-syn}} & \multicolumn{2}{c}{\textbf{Efficiency}} \\
        \textbf{} & \textbf{PSNR$\uparrow$} & \textbf{SSIM$\uparrow$} & \textbf{LPIPS$\downarrow$} & \textbf{PSNR$\uparrow$} & \textbf{SSIM$\uparrow$} & \textbf{PSNR$\uparrow$} & \textbf{SSIM$\uparrow$} & \textbf{\# of P (M)$\downarrow$} & \textbf{FLOPs (G)$\downarrow$} \\
        \hline
        LIME~\cite{Guo2017lime} & 16.67 & 0.560 & 0.368 & 15.24 & 0.470 & 17.63 & 0.787 & - & - \\
        Retinex-Net \cite{Chen2018Retinex} & 16.77 & 0.425 & 0.474 & 18.37 & 0.723 & 17.14 & 0.756 & 0.555 & 587.470 \\
        KinD \cite{zhang2019kind} & 17.65 & 0.771 & {0.175} & 14.74 & 0.641 & 17.28 & 0.758 & 8.160 & 574.950 \\
        EnGAN \cite{Jiang2021EnlightenGAN} & 17.54 & 0.664 & 0.326 & 18.23 & 0.617 & 16.49 & 0.771 & 114.350 & 223.430 \\
        Zero-DCE \cite{guo2020zerodce} & 14.86 & 0.562 & 0.335 & 14.32 & 0.511 & 17.76 & 0.814 & 0.075 & 4.830 \\
        RUAS \cite{liu2021ruas} & 18.23 & 0.720 & 0.270 & 15.33 & 0.488 & 13.76 & 0.634 & \textbf{0.003} & \textbf{0.830} \\
        IAT \cite{Cui_2022_IAT} & \underline{23.38} & 0.809 & 0.210 & \underline{23.50} &  {0.824} & 15.37 & 0.710 & 0.091 & 5.271 \\
        PairLIE \cite{fu2023pairlie} & 19.56 & 0.730 & {0.248} & 19.89 & 0.778 & 19.07 & 0.794 & 0.342 & 81.838 \\
        FLIGHT-Net \cite{ozcan2023flight} & \textbf{24.96} & \textbf{0.850} & \textbf{0.134} & {21.71} & {0.834} & \textbf{24.92} & \textbf{0.930} & 0.025 & 3.395 \\
        LLFormer~\cite{wang2023llformer} & 23.65 & 0.816& 0.169 & \textbf{27.75} &\textbf{0.860} & 17.16 & 0.784 & 24.55 & 39.05 \\
        DDNet~\cite{qu2024ddnet} & 21.82 & 0.798 & 0.186 & {23.02} & {0.834} & \underline{24.63} & {0.917} & 5.390 & 111.47 \\
        LYT-Net~\cite{brateanu2025lyt} & 22.38 & 0.826 & 0.134 & 20.97 & 0.840 & 23.50 & 0.914 & 0.045 & 8.037\\
        FLOL+~\cite{benito2025flol} & 21.07 & 0.812 & 0.195 & 22.15 & {0.846} & 18.56 & 0.862 & 0.095 & 3.743 \\
        CPGA-Net \cite{weng2024lightweight} & 20.94 & 0.748 & 0.260 & 20.79 & 0.759 & 20.68 & 0.833 & 0.025 & 6.030 \\
        CPGA-DIA~\cite{weng2024dia} & 20.37 & 0.760  & 0.280 & 22.18 & 0.794 & 18.22 & 0.799 & 0.065 & 15.520\\
        \hline
        CPGA-Net+ & 22.53 & {0.812} & 0.205 & 20.90 & 0.800 & {23.07} & {0.907} & 0.060 & 9.356 \\
        CPGA-Net+ (BDSF) & 22.53 & {0.812} & 0.205 & 20.90 & 0.800 & {23.07} & 0.907 & \underline{0.020} & \underline{2.141} \\
        CPGA-Net++ & 22.24 & \underline{0.835} & \underline{0.136} & 21.29 & \underline{0.850} & {24.31} & \underline{0.920} & 0.062 & 13.285 \\
        \hline
    \end{tabular}
\end{table*}

\begin{table*}[!t]
    \centering
    \caption{The image quality comparison on unpaired data \cite{Guo2017lime, Ma2015MEF, Wang2013NPE, Vonikakis2018VV, Lee2013DICM} in terms of the NIQE metric, where lower values generally indicate better performance. We represent the first and second ranks with \textbf{bold} and \underline{underlined}, respectively. For the learning methods, T indicates the traditional approach, U indicates unsupervised learning, and S indicates supervised learning. BDSF means Block Design Simplification for our approach.}
    \label{table.comp_unpair}
        \begin{tabular}{lccccccccc}
            \hline
            & & \multicolumn{5}{c}{\textbf{Datasets}} \\
            \textbf{Original Image and Method} & \textbf{Types} & \textbf{MEF} & \textbf{LIME} & \textbf{NPE} & \textbf{VV} & \textbf{DICM} & \textbf{Avg} \\
            \hline
            Low-light Image & N/A & 4.2650 & 4.4380 & 4.3190 & 3.5350 & 4.2550 & 4.1624 \\
            \hline
            NPE \cite{Wang2013NPE} & T & 3.5240 & 3.9048 & 3.9530 & 2.5240 & 3.7600 & 3.5332 \\
            LIME \cite{Guo2017lime} & T & 3.7200 & 4.1550 & 4.2680 & 2.4890 & 3.8460 & 3.6956 \\
            \hline
            EnlightenGAN \cite{zhang2019kind} & U & \textbf{3.2320} & 3.7190 & 4.1130 & 2.5810 & 3.5700 & 3.4430 \\
            Zero-DCE \cite{guo2020zerodce} & U & 4.0410 & 3.7890 & 3.5041 & 2.7526 & 3.1018 & 3.4377 \\
            RUAS \cite{liu2021ruas} & U & 4.1403 & 4.2900 & 4.8713 & 3.5086 & 4.5417 & 4.2704 \\
            PairLIE \cite{fu2023pairlie} & U & 4.0862 & 4.3113 & 4.0890 & 3.1595 & 3.2422 & 3.7776 \\
            CPGA-Net+ \cite{weng2024lightweight} & U & 3.5950 & 3.2575 & 3.4438 & 2.8820 & 3.0350 & 3.2427 \\
            \hline
            KinD \cite{zhang2019kind} & S & 3.8830 & {3.3430} & 3.7240 & 2.3208 & 2.9888 & 3.2519 \\
            IAT \cite{Cui_2022_IAT} & S & 3.6188 & 4.1722 & {3.2890} & 2.5270 & 3.0325 & 3.3279 \\
            FLIGHT-Net \cite{ozcan2023flight} & S & 3.5491 & 3.7049 & 3.3311 & 2.9435 & 2.8979 & 3.2853 \\
            LLFormer~\cite{wang2023llformer} & S & 3.3588 &	3.6798	& 3.2570 &	2.2712	& 2.9407 & 3.1015 \\
            DDNet \cite{qu2024ddnet} & S & {3.2734} & 3.4329 & 3.1135 & 2.0223 & {2.6409} & 2.8970 \\
            LYT-Net~\cite{brateanu2025lyt} & S & 3.5152 & 3.3929 & \textbf{2.9690} & 2.3812 & 2.8225 & 3.0162 \\
            FLOL+~\cite{benito2025flol} & S & \underline{3.2409} & 3.6036 & \underline{2.9710} & 2.5315 & \textbf{2.4124} & 2.9519 \\
            CPGA-Net \cite{weng2024lightweight} & S & 3.8698 & 3.7068 & 3.5476 & 2.2641 & {2.6934} & 3.2163 \\
            CPGA-DIA \cite{weng2024dia} & S & 3.5880 & 3.5570 & 3.1650 & 2.0930 & {2.6300} & 3.0006 \\
            \hline
            CPGA-Net+ & S & {3.4968} & \underline{3.0626} & {3.0886} & \textbf{1.9133} & {2.8282} & \underline{2.8779} \\
            CPGA-Net+ (BDSF) & S & 3.4969 & {3.0655} & {3.0881} & \underline{1.9136} & {2.8268} & {2.8782} \\
            CPGA-Net++ & S & 3.3825 & \textbf{2.8646} & 3.0434 & 1.9302 & \underline{2.5241} & \textbf{2.7490}\\ 
            \hline
        \end{tabular}
\end{table*}

\subsection{High-Level Vision Task}
In this section, we address the challenge of objection detection in low-light environments by utilizing a joint training approach of YOLOv9s~\cite{wang2025yolov9} with the SOTA approaches of LLIE on the ExDark dataset~\cite{loh2019exdark}, as illustrated in Table~\ref{table.detection}. Our approach improves the mean Average Precision (mAP) by 0.075 compared to the baseline. All the LLIE methods listed here can improve object detection performance, among which the Zero-DCE~\cite{guo2020zerodce} and our proposed method are the best. The results show that our proposed method can improve not just human perception but machine perception as well.

\begin{table}[!ht]
    \centering
    \caption{Comparison of performance metrics between YOLOv9s with CPGA-Net+ and other SOTA methods on the ExDark dataset~\cite{loh2019exdark}.}
    \label{table.detection}
    \begin{adjustbox}{width=\linewidth,center}
        \begin{tabular}{lcccccc}
            \hline
            \textbf{Method} & \textbf{Precision$\uparrow$} & \textbf{Recall$\uparrow$} & \textbf{mAP@.5$\uparrow$} & \textbf{mAP@.5:.95$\uparrow$} \\
            \hline
            YOLOv9s \cite{wang2025yolov9} & 0.745 & 0.562 & 0.639 & 0.419 \\
            YOLOv9s + Zero-DCE \cite{guo2020zerodce} & \textbf{0.801} & \textbf{0.616} & \textbf{0.714} & \underline{0.470} \\
            YOLOv9s + IAT \cite{Cui_2022_IAT} & 0.725 & 0.600 & \underline{0.675} & 0.445 \\
            YOLOv9s + CPGA-Net+ & \underline{0.790} & \underline{0.601} & \textbf{0.714} & \textbf{0.471} \\
            \hline
        \end{tabular}
    \end{adjustbox}
\end{table}

\section{Ablation Study}
In this section, we analyze the effectiveness of each systematic module and training technique, including the systematic design and integration, the number of Channel-Prior blocks, and loss functions.

\subsection{Systematic Design and Integration}
As shown in Table~\ref{table.ablation_design}, our method effectively fuses the gamma correction from the global branch to the local branch, resulting in improved overall performance and demonstrating the strength of our approach. By grounding the attention mechanism in gamma correction, we ensure that the enhancement process remains aligned with the non-linearities inherent in the imaging process and human perception.

On the other hand, we received the surprising results for our lightweight (BDSF) and stronger pipeline (CPGA-Net++) designs, as shown in Tables~\ref{table.ablation_design} and \ref{tab:ablation_cpga}. The BDSF strongly utilizes our model's characteristics based on the fact that global processing dominates and suppresses the local processing, using an antagonism to maximize the power of global processing, leading to an ultimate and simple pruning to reduce the computational cost of local processing. In contrast, CPGA-Net++ employed IAAF+ to encourage both branches to operate effectively, resulting in improved performance. However, this approach diminishes the auxiliary function of local processing, making it impossible to achieve performance levels comparable to BDSF for the efficient version via pruning.

\begin{table}[t]
    \centering
    \caption{Ablation study of systematic design. L-G denotes our design of plug-in attention from global to local processing, utilizing the CPGA block to bridge the gap between local and global branches. (f) shares the weights from (e) but performs inference via global processing only, sharing the same design as (c). (e)* denotes using the weights obtained from (e).}
    \label{table.ablation_design}
    \begin{adjustbox}{width=\linewidth,center}
        \begin{tabular}{lccccccccc}
            \hline
            & \multicolumn{3}{c}{\textbf{Network Design}} & \textbf{Training} & \multicolumn{3}{c}{\textbf{LOLv1}} & \multicolumn{2}{c}{\textbf{Efficiency}}\\
            &\textbf{Local} & \textbf{L-G} & \textbf{Global} & \textbf{BDSF} & \textbf{PSNR$\uparrow$} & \textbf{SSIM$\uparrow$} & \textbf{LPIPS$\downarrow$} & \textbf{\# of P. (M)$\downarrow$} & \textbf{FLOPs (G)$\downarrow$} \\
            \hline
            (a) & \checkmark & & && 18.36 & 0.743 & 0.297 & 0.030 & 4.78 \\
            (b) & \checkmark & \checkmark & && 20.82 & 0.782 & 0.254 & 0.056 & 8.200 \\
            (c) & &  & \checkmark && 22.08 & 0.810 & 0.188 & 0.020 & 2.141 \\
            (d) & \checkmark & & \checkmark && 20.87 & 0.803 & 0.205 & 0.050 & 6.929 \\
            (e) & \checkmark & \checkmark & \checkmark && 22.53 & 0.812 & 0.205 & 0.060 & 9.356 \\
            (f) & & & \checkmark  & (e)* & 22.53 & 0.812 & 0.205 & 0.020 & 2.141 \\
            \hline
        \end{tabular}
    \end{adjustbox}
\end{table}

\begin{table*}[ht]
    \centering
    \caption{Ablation study of structural functions for CPGA-Net++. The type indicates the fusion input source: Homo refers to homogeneous fusion using \( R^\gamma \), while Hetero refers to heterogeneous fusion using \( L^\gamma \). For the efficient version comparison, (b) is CPGA-Net+ (BDSF), and (i) is the finetune of CPGA-Net++ with the same structural pruning as BDSF, as CPGA-Net++ is not possible to directly prune while the characteristic of near-identity mapping is no longer maintained; therefore, we added a * beside the BDSF.}
    \begin{adjustbox}{width=\linewidth,center}
    \begin{tabular}{c|cc|ccc|c|ccc|cc}
        \hline
        & \multicolumn{2}{c|}{\textbf{Block}} & \multicolumn{3}{c|}{\textbf{IAAF}} & \textbf{Efficient} & \multicolumn{3}{c|}{\textbf{IQA} } & \multicolumn{2}{c}{\textbf{Efficiency}}\\
        & \textbf{Resblock} & \textbf{ConvNeXT} & \textbf{Type} & \textbf{WFS} & \textbf{IRS} & &
        \textbf{PSNR↑} & \textbf{SSIM↑} & \textbf{LPIPS↓} & \textbf{\# of P. (M)} & \textbf{FLOPs (G)} \\
        \hline
        (a) & \checkmark  &  & Homo &  & & & 22.53 & 0.812 & 0.205 & 0.060 & 9.356 \\
        (b) & \checkmark &  & Homo &  & & BDSF & 22.53 & 0.812 & 0.205 & 0.020 & 2.141 \\
        (c) & \checkmark &  & Hetero  &  & &  & 22.42 & 0.810 & 0.206 & 0.060 & 9.356 \\
        (d) & \checkmark &  & Hetero  & \checkmark &  & & 21.94 & 0.812 & 0.194 & 0.060 & 9.230 \\
        (e) & \checkmark &  & Hetero &  & \checkmark & & 21.87 & 0.802 & 0.211 & 0.089 & 22.793 \\
        (f) & \checkmark &  & Hetero & \checkmark & \checkmark & & 22.89 & 0.814 & 0.182 & 0.089 & 22.793 \\
        (g) & \checkmark &  & Homo & \checkmark & \checkmark & & 21.34 & 0.799 & 0.231 & 0.089 & 22.793 \\
        (h) &  & \checkmark & Hetero & \checkmark & \checkmark & & 22.24 & 0.835 & 0.136 & 0.062 & 13.302 \\
        (i) &  & \checkmark & Hetero & \checkmark & \checkmark & BDSF* & 21.28 & 0.800 & 0.220 & 0.023 & 3.696 \\
        \hline
    \end{tabular}
    
    \end{adjustbox}
    \label{tab:ablation_cpga}
\end{table*}

\subsection{The Number of Channel-Prior Blocks}
We explore how varying the number of CP blocks affects the model's capacity to enhance image quality. The results, summarized in Table~\ref{table.ablation_cp}, show that increasing the number of CP blocks leads to an improvement from 0 to 2 blocks but show no significant changes from 2 to 4 blocks. However, the number of parameters and computational cost (FLOPs) increase with more CP blocks, introducing greater computational demands. Therefore, the optimal number of CP blocks should balance performance gains with resource efficiency. We selected 2 CP blocks for our final approach to achieve a lightweight and efficient design.
\begin{table}[!t]
    \centering
    \caption{Ablation study of the number of CP blocks. \(N=2\) is the default setting of our approach.}
    \label{table.ablation_cp}
    \begin{tabular}{lccccc}
        \hline
        \textbf{N} & \textbf{PSNR$\uparrow$} & \textbf{SSIM$\uparrow$} & \textbf{LPIPS$\downarrow$} & \textbf{\# of P. (M)$\downarrow$} & \textbf{FLOPs (G)$\downarrow$} \\
        \hline
        0 & 20.56 & 0.754 & 0.250 & 0.034 & 4.243 \\
        2 & 22.53 & 0.812 & 0.205 & 0.060 & 9.356 \\
        4 & 21.93 & 0.805 & 0.215 & 0.087 & 14.503 \\
        \hline
    \end{tabular}
\end{table}

\subsection{Loss Functions}
This section examines the impact of various loss function combinations on the model's performance. We tested L1 loss, Perceptual loss, HDR L1 loss, and SSIM loss, with the results summarized in Table~\ref{table.ablationloss}. Using the default settings as in CPGA-Net, the combination of L1 and Perceptual losses performs well, yielding a PSNR improvement of 2.82 dB and an SSIM increase of 0.024. The HDR L1 loss significantly enhances all three metrics, with PSNR increasing by 2.86 dB, SSIM by 0.041, and LPIPS decreasing by 0.085. While SSIM loss improves its specific metric with an SSIM boost of 0.031, it is less effective in enhancing PSNR. Ultimately, combining all these losses results in the best overall performance for supervision, which improves PSNR by 4.04 dB, SSIM by 0.083, and LPIPS by 0.118.

\begin{table}[t]
    \centering
    \caption{Ablation study of loss functions.}
    \label{table.ablationloss}
        \begin{tabular}{lccccccc}
            \hline
            \textbf{} & \textbf{L1} & \textbf{Per} & \textbf{HDR L1} & \textbf{SSIM} & \textbf{PSNR$\uparrow$} & \textbf{SSIM$\uparrow$} & \textbf{LPIPS$\downarrow$} \\
            \hline
            (a) & \checkmark &  &  &  & 18.49 & 0.729 & 0.323 \\
            (b) & \checkmark & \checkmark &  &  & 21.31& 0.753 & 0.250 \\
            (c) & \checkmark & \checkmark & \checkmark &  & 21.35 & 0.770& 0.238 \\
            (d) & \checkmark & \checkmark &  & \checkmark & 20.41  & 0.760  & 0.243  \\
            (e) & \checkmark & \checkmark & \checkmark & \checkmark & 22.53 & 0.812 & 0.205  \\
            \hline
        \end{tabular}
\end{table}

\section{Limitation}
\begin{figure}[t]
    \centering
    \includegraphics[width=\linewidth]{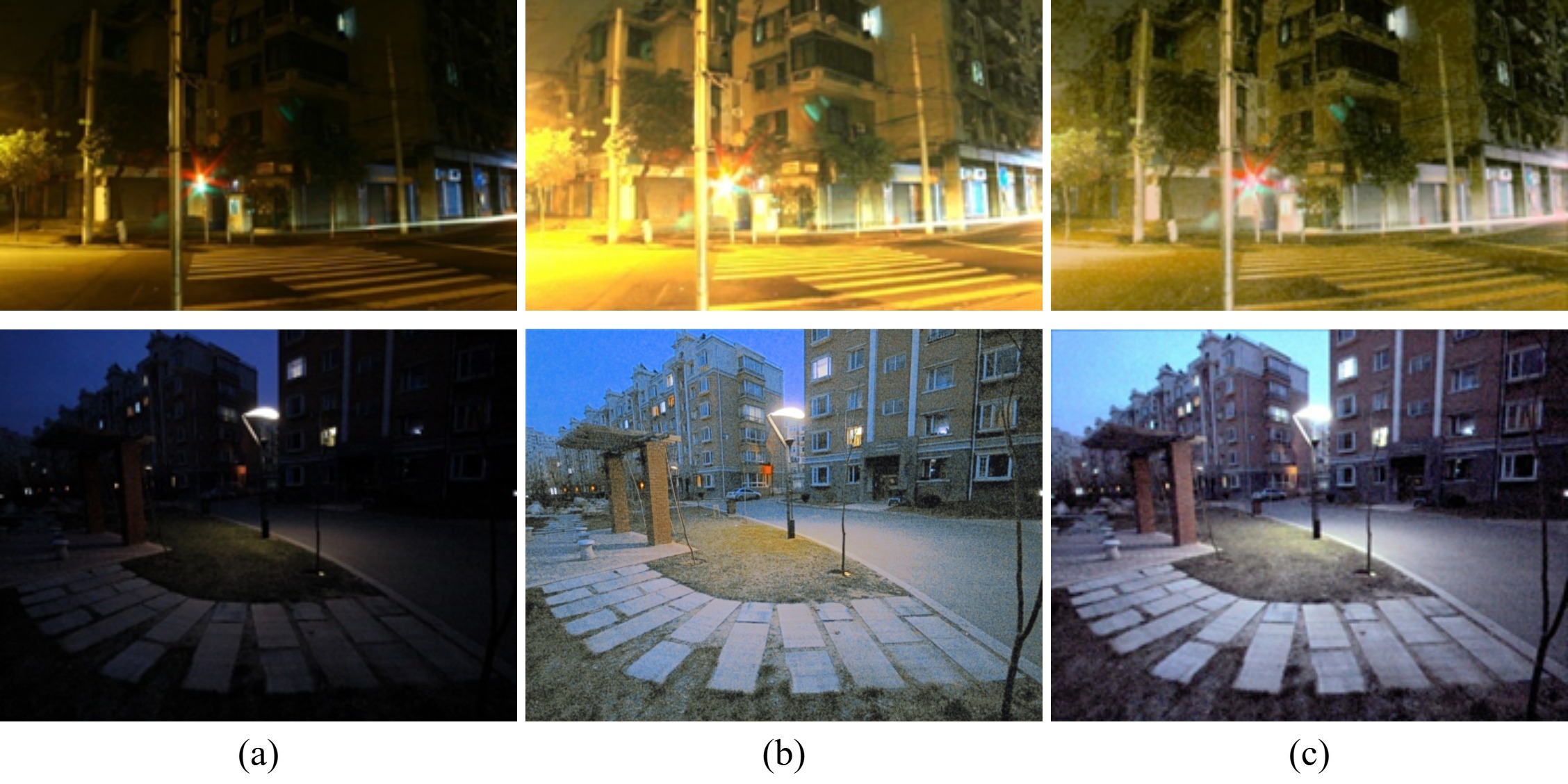}
    \caption{Visualization of extreme low-light scenarios~\cite{Guo2017lime, Lee2013DICM}. (a) Original images; (b) Supervised CPGA-Net+; (c) Unsupervised CPGA-Net+. While the enhancement tone appears satisfactory, both methods exhibit defects such as pepper-and-salt noise, resulting in a grainy texture.}
    \label{fig.limit}
\end{figure}

While our proposed method shows notable improvements in efficiency and performance, a key limitation warrants consideration. As discussed in the methodology section, our approach benefits from the guidance of Channel-Priors and Gamma Correction, which enhances contrast and visual perception. However, gamma correction for brightening with a small value of estimated gamma can lead to defects or distortions in extremely low-light scenarios, even while maintaining realistic exposure. Fig.~\ref{fig.limit} shows an example of such a limitation. Addressing this issue will require further refinement for brightening and denoising, which we aim to pursue in future work.

\section{Conclusion}
This work looks deeper at CPGA-Net, utilizing it as an attention mechanism grounded in theoretical formulas. We propose a stacked and modularized attention module to focus on image details and integrate gamma correction into the local branch, creating a Plug-in Attention module for each CP block. Moreover, we reevaluate the systematic design and propose two expanded applications with structural pruning and maximizing potential through theoretical and practical analysis. This enhancement makes our approach lightweight yet SOTA in performance, maintaining strong efficiency and stable operation across devices with limited computational resources. In the future, we are striving to improve the unsupervised learning of CPGA-Net+ and integrate our approach into HDR imaging and exposure fusion, improving detail preservation in both bright and dark areas by leveraging brightness sensitivity through prior knowledge, such as channel and gamma-correction priors. This will enhance the output's dynamic range and overall fidelity in real-world applications.


%



\section*{Acknowledgment}
This work was partly supported by the Ministry of Science and Technology, Taiwan, under Grant NSTC 113-2221-E-033-055.

\clearpage

\appendices

\begin{figure*}
    \centering
    \includegraphics[width=\textwidth]{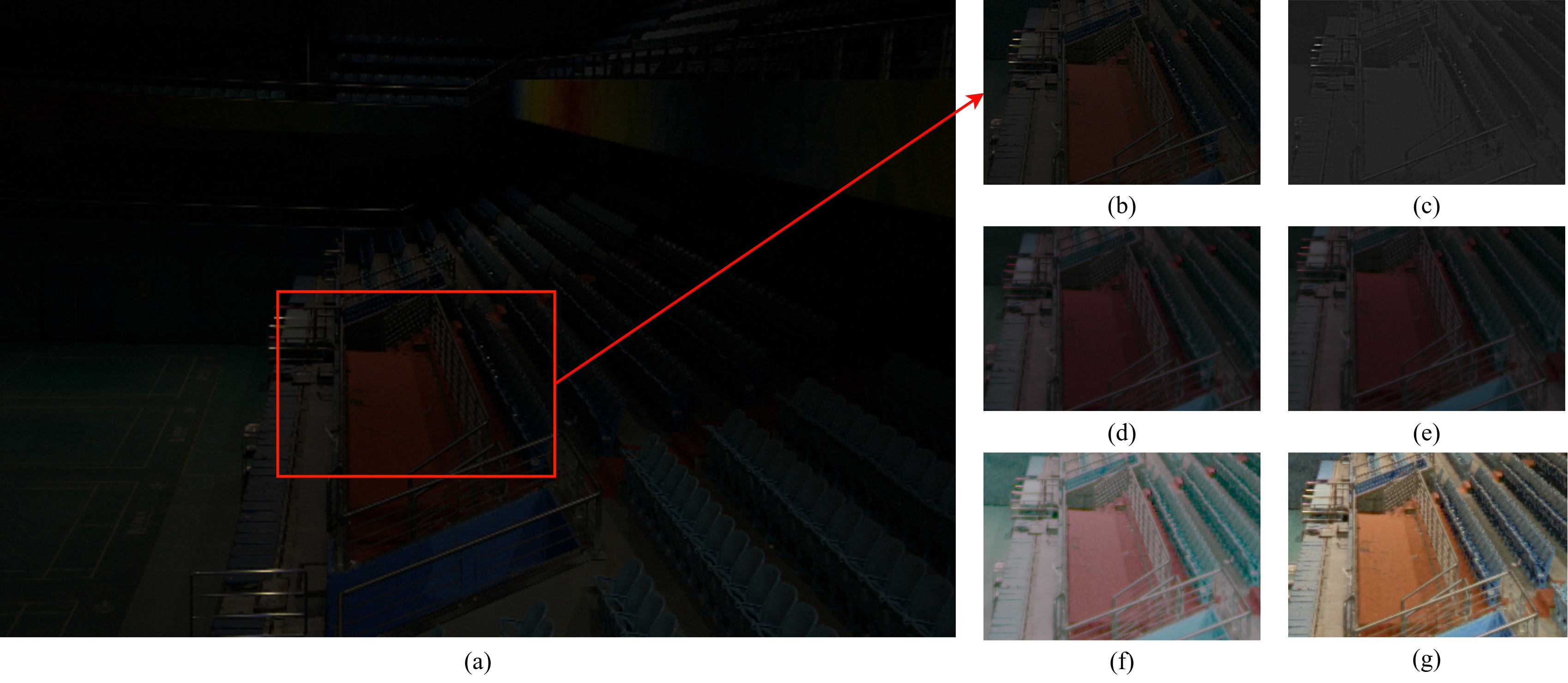}
    \caption{Visualization of extracted features~\cite{weng2024lightweight}. (a) Low-light image \(L\); (b) A portion of \(L\); (c) \(\tilde{t}\) (contribution proportion from  \(L\)); (d) An estimated image of \(\tilde{A}\); (e) Reconstructed low-light image \(R\); (f) Gamma-corrected image \(R^\gamma\); (g) Enhanced image \(\hat{R}\).}
    \label{fig.feature_maps}
\end{figure*}

\section{Revealing the Explainable Mechanisms Behind Degradation}\label{subsec.explainable}
Our method follows a rule-based learning strategy, leveraging the ATSM and the Retinex theory to perform image processing based on physical models. The ATSM simulates the scattering and absorption of light in uneven media, while the Retinex theory, inspired by retinal imaging, models the human visual system. Although these methods provide solid theoretical support from both physical and physiological perspectives, challenges remain in the interpretability of neural networks. While applying the models to deep learning, it is challenging to explain their decision-making mechanisms due to the highly nonlinear nature of deep learning models. Similarly, these aspects are difficult to define. They cannot be easily explained through experience or a basic understanding of the equations derived from the generated features in the neural network. To improve the interpretability of our approach, we use the traditional image degradation model, which offers a more explicit theoretical framework for the deep learning process. This effectively enhances the transparency of the entire neural network, opening the “black box” of image enhancement.

The fundamental image degradation model in matrix form~\cite{gonzalez2017digital} can be defined as:
\begin{equation}
    \textbf{G} = \textbf{HF} + \textbf{N}
\end{equation}
where $\textbf{G}$ is the image after degradation, $\textbf{H}$ is the unknown degradation kernel, $\textbf{N}$ denotes additive noise, and $\textbf{F}$ is the image before degradation. The restoration process is aimed at estimating $\textbf{F}$ by:
\begin{equation}
    \hat{\textbf{F}} = \textbf{H}^{-1} (\textbf{G} - \textbf{N}) = \textbf{H}^{-1} \textbf{R}
\end{equation}
Here, the estimated and enhanced image $\hat{R}$ corresponds to $\hat{\textbf{F}}$. Based on the visual observation results from~\cite{weng2024lightweight}, the local branch can be seen as a reconstruction and denoising process for the noise-free or ground-truth reconstructed image $R^{\text{GT}}$, which is equivalent to $\textbf{R} = \textbf{G} - \textbf{N}$, and the global branch can be viewed as an information extraction process for $\textbf{H}^{-1}\textbf{R}$ to lighten the image for a more natural appearance. The characteristic of $\textbf{H}^{-1}$ is extracted by the IAAF module represented by Eq.~\eqref{eq.IAAF} in the main text. Some relevant images under the current discussion are shown in Fig.~\ref{fig.feature_maps}.

We attempt to further verify our hypotheses for the local branch through least-squares optimization~\cite{boyd2004convex}. First, we make the following simple assumptions: we assume that \(\hat{A}\) and $L$ can be represented as the noise-free reconstructed image $R^{\text{GT}}$ with additional composite noises that consist of three-channel mixing factors, lighting changes, and inherent thermal noises, which are particularly noticeable in low-light environments, as depicted below:
\begin{equation}
    \tilde{A} = R^{\text{GT}} + N_A
    \label{eq.A_N}
\end{equation}
\begin{equation}    
    L = R^{\text{GT}} + N_L
    \label{eq.L_N}
\end{equation}
where \( L \) is the low-light image, \( N_A \) and \( N_L \) denote the noise components associated with \( \tilde{A} \) and \( L \), respectively. While estimating \( N_A \), we assume that \( N_L \) remains fixed once the low-light image \( L \) is given. Then, we can define the following residual error $r(\tilde{t}, N_A)$ by combining Eq.~\eqref{eq.enhance_atsm} in the main text, Eq.~\eqref{eq.A_N}, and Eq.~\eqref{eq.L_N}:

\begin{equation}
    r(\tilde{t}, N_A) = R - R^{\text{GT}} = (1 - \tilde{t}) N_A + \tilde{t} N_L
\end{equation}
where \( R \) is the reconstructed image, \( \tilde{t} \) is a positive parameter used to adjust the reconstruction process. We aim to find \( N_A \) and \( \tilde{t} \) such that the residual error is minimized, allowing \( R \) to approach \( R^{\text{GT}} \).

We further define a cost function $C(\tilde{t}, N_A)$ for the least-squares optimization process:
\begin{equation}
    C(\tilde{t}, N_A) = \frac{1}{2} \left[ r(\tilde{t}, N_A) \right]^2
\end{equation}

Next, to find the stationary points in the optimization process, we conduct partial differentiation of $C(\tilde{t}, N_A)$ with respect to $\tilde{t}$ and $N_A$:

\begin{equation}
\frac{\partial C}{\partial \tilde{t}} = r(\tilde{t}, N_A) \cdot \left({N_L - N_A}\right) = 0
\end{equation}
\begin{equation}
    \frac{\partial C}{\partial N_A} = r(\tilde{t}, N_A) \cdot \left(1 -\tilde{t}\right) = 0
\end{equation}

\noindent\textbf{Case I:} $r(\tilde{t}, N_A) \neq 0$: We have
\begin{equation}
    (N_A - N_L) = 0 \quad \text{and} \quad \tilde{t} = 1
\end{equation}
which leads to the solution $N_A = N_L$, resulting in:
\begin{equation}
    r_{\text{min}}(\tilde{t}, N_A) = N_L
\end{equation}

However, it results in $R = L$ from Eq.~\eqref{eq.enhance_atsm} in main text, meaning that no restoration effect is involved at all (identity mapping from the input $L$ to the output $R$).

\noindent\textbf{Case II:} $r(\tilde{t}, N_A) = 0$: We have

\begin{equation}
    R - R^{\text{GT}} = 0 \implies R = R^{\text{GT}}
    \label{eq.case2}
\end{equation}
and
\begin{equation}
    (1 - \tilde{t}) N_A + \tilde{t} N_L = 0 \implies N_L = \left(1 - \frac{1}{\tilde{t}} \right) N_A, \quad \tilde{t} \neq 0
    \label{eq.NA_NL}
\end{equation}
Therefore, the optimal solution exists when Eq.~\eqref{eq.NA_NL} holds.

The results show that the enhancement follows the degradation formula, with the ATSM simulating the network's processing of illumination, reflection, and noise, supporting the rationality of neural networks in LLIE. This enhances the model’s transparency and reveals its interpretability mechanism, offering deeper insights into how neural networks function in image enhancement tasks.

From a learning mechanism perspective, deep learning exhibits significant similarities to human cognition, emphasizing that the most prominent features carry essential information rather than relying on intricate details, with \( \tilde{A} \) for detail restoration representing more critical information than \( \tilde{t} \) for contrast and saturation. By modularizing and extending the understanding through attention mechanisms, we effectively capture the relationships between local and global information, enhancing image quality in low-light conditions and yielding better results across various complex scenarios. This method underscores the potential for integrating traditional theory with deep learning models, offering valuable insights for future technological advancements and opening new research opportunities in image enhancement.




\clearpage
\bibliographystyle{IEEEtran}
\bibliography{bibtex/bib/IEEEabrv, ref}
\end{document}